\documentclass[final,3p,times,twocolumn,authoryear]{elsarticle}

\usepackage{amssymb}
\usepackage{amsmath}

\usepackage{tabularx}
\usepackage{booktabs}
\usepackage{pifont}
\usepackage{graphicx}
\usepackage{xcolor}
\usepackage{booktabs,multirow}
\usepackage{makecell}
\usepackage{float}
\usepackage{stfloats}
\usepackage{arydshln}
\usepackage{xcolor}
\usepackage{longtable}
\usepackage{array} 
\definecolor{myblue}{rgb}{0.21, 0.49, 0.74}
\usepackage{hyperref}
\hypersetup{
    colorlinks=true
}

\usepackage{colortbl}
\usepackage{caption}

\definecolor{darkgreen}{rgb}{0.45, 0.75, 0.45} 
\definecolor{green}{rgb}{0.7, 0.9, 0.6} 
\definecolor{orange}{rgb}{1.0, 0.8, 0.6} 
\definecolor{darkred}{rgb}{1.0, 0.6, 0.6} 

\definecolor{lightblue}{RGB}{185,235,250}
\definecolor{lightpurple}{RGB}{240,205,235}
\definecolor{lightgreen}{RGB}{225,245,190}

\journal{Neural Networks}

\begin{document}

\begin{frontmatter}



\title{CAPS: Clinically-Verifiable Anatomy-Prioritized Synthesis for Missing MRI Imputation in Brain Tumor Segmentation}


\author{Cecilia Diana-Albelda$^{a, \dagger, *}$, Arthur Longuefosse$^{b, \dagger}$, Mario Rubén Paz Campos$^{c}$, Álvaro García-Martín$^{a}$, Jesus Bescos$^{a}$ \\[0.4cm] 
\footnotesize $^{\dagger}$ Equal contribution. \quad $^{*}$ Corresponding author: \href{mailto:cecilia.diana.albelda@gmail.com}{cecilia.diana.albelda@gmail.com}
}

\affiliation{organization={$^{a}$ Universidad Autónoma de Madrid, Department of Electronic Technology and Communications},
            city={Madrid},
            country={Spain}}

\affiliation{organization={$^{b}$ RIKEN Center for Integrative Medical Sciences, Medical Data Deep Learning Team},
            city={Tokyo},
            country={Japan}}

\affiliation{organization={$^{c}$ Hospital Universitario La Princesa, Diagnostic Radiology Department},
            city={Madrid},
            country={Spain}}

\begin{abstract}
Multi-parametric MRI is the gold standard for neuro-oncology, yet acquiring complete 4-modality protocols may be unfeasible in real-world clinical practice. When handling missing modalities, existing automated systems force a critical compromise: robust segmenters operate as uninterpretable "black boxes" that bypass image generation, while standard synthesis models often produce blurry reconstructions lacking diagnostic reliability. To address this issue, we propose CAPS, a Clinically-Verifiable Anatomy-Prioritized Synthesis framework that reconciles computational precision with clinical transparency. Our multi-modal generative architecture is driven by a hybrid objective function ($\mathcal{L}_1 + \mathcal{L}_{AFP} + \mathcal{L}_{DualDice}$) that explicitly enforces the preservation of deep anatomical features and strict topological boundaries, rather than relying solely on pixel-wise optimization. Evaluated comprehensively on the BraTS Adult Glioma dataset, the framework achieves superior synthesis fidelity across intensity-based metrics, and demonstrates robust downstream segmentation performance ($82.38\%$ mean Dice), successfully restoring verifiable visual context. Furthermore, the framework exhibits exceptional zero-shot cross-domain generalization, maintaining high structural fidelity across completely unseen Meningioma, Pediatric, and Sub-Saharan African cohorts without domain-specific adaptation. Finally, a blinded radiological assessment confirms that our anatomy-aware approach significantly enhances both perceived image quality and clinical utility, providing clinicians with the reliable visual evidence required to trust downstream automated workflows.
\end{abstract}




\end{frontmatter}



\section{Introduction}

Gliomas are among the most prevalent primary Central Nervous System (CNS) tumors \cite{weller2015glioma}, with high-grade gliomas being particularly aggressive due to their highly infiltrative growth and low survival rates \cite{guo2023clinical}. The gold standard for the diagnosis, treatment planning, and post-operative monitoring of these tumors is multi-parametric Magnetic Resonance Imaging (mp-MRI). A standard neuro-oncological protocol typically comprises four scan modalities: Native T1-weighted (T1n), Post-contrast T1-weighted (T1c), T2-weighted (T2w), and T2 Fluid-Attenuated Inversion Recovery (T2f) \cite{diana2025review}. Each modality provides unique and complementary tissue contrasts that are critical for accurate tumor characterization. For instance, the active, proliferating tumor core and neoangiogenesis are primarily visible in the gadolinium-enhanced T1c sequence, while the surrounding peritumoral edema and non-enhancing infiltrative regions are best delineated using T2f. However, acquiring the complete 4-modality protocol in real-world clinical settings is sometimes unfeasible, as scans may be incomplete due to time constraints, patient motion artifacts, financial limitations, or contraindications to gadolinium-based contrast agents \cite{li2024brain}. The absence of even a single modality introduces a twofold clinical and algorithmic challenge. Clinically, it deprives neuroradiologists of vital anatomical information, potentially leading to suboptimal surgical planning and, ultimately, negatively impacting patient survival and recurrence rates. Algorithmically, state-of-the-art automated segmentation systems, which are increasingly integrated to assist clinical workflows, are typically designed to process the full set of four input channels, thereby experiencing severe performance degradation or failure when a modality is missing or zero-padded \cite{sharma2019missing}.

To mitigate this problem, two main strategies have emerged in the literature. The first one focuses on designing robust segmentation frameworks that can predict tumor boundaries despite incomplete inputs \cite{zhang2022mmformer, pipoli2025fuse}. While these systems achieve remarkable accuracy, they present a significant challenge in clinical practice: they output a tumor mask without restoring the missing anatomical image, leaving the clinician without the visual evidence needed to corroborate the automated delineation. The second strategy relies on image synthesis to impute the missing modality \cite{diana2025multi}. However, traditional synthesis methods primarily optimize standard pixel-wise reconstruction losses, such as L1, which frequently results in blurry outputs that miss important structural details \cite{longuefosse2025anatomical}, thereby restricting their practical utility for the clinicians.

\begin{figure}[ht!]
\centering
\includegraphics[width=0.40\textwidth]{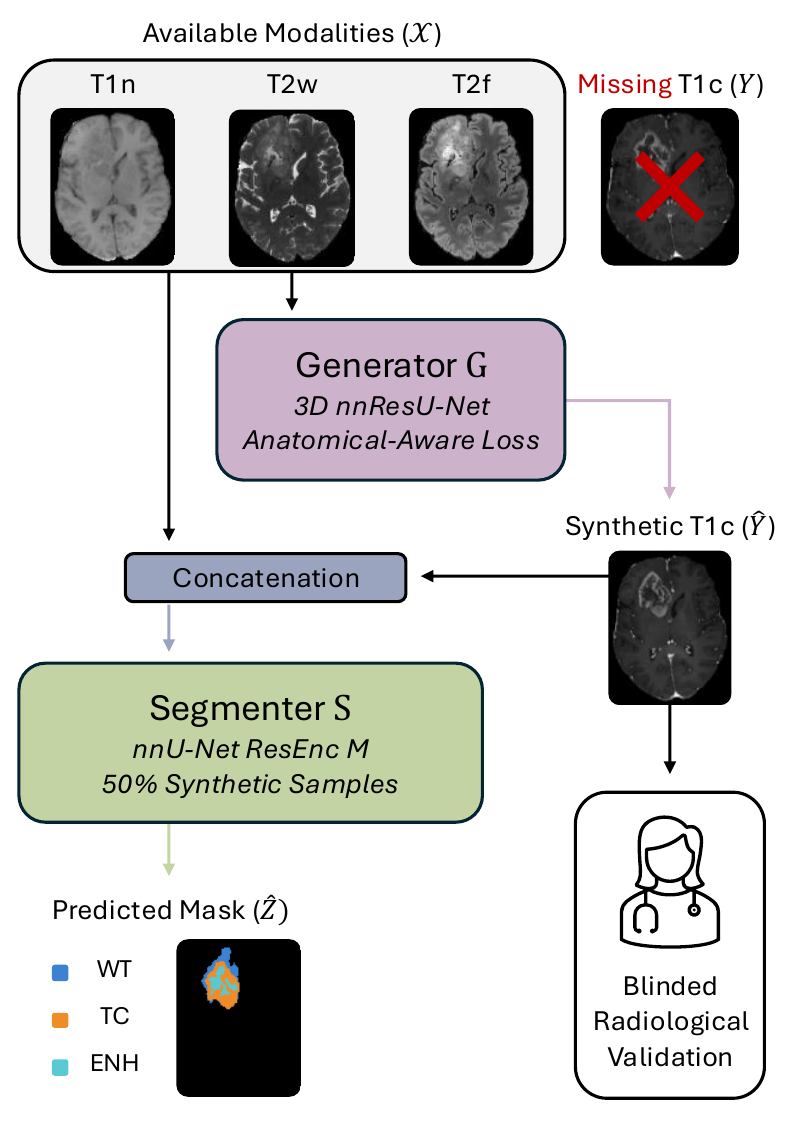}
\caption{\textbf{Overview of CAPS.} Given three available MRI modalities ($\mathcal{X}$: T1n, T2w, T2f), the generator $G$ synthesizes the missing sequence ($\hat{Y}$, T1c in this example). The synthesized volume is then concatenated with the available modalities and fed into the clinical segmenter $S$, which produces the predicted tumor mask ($\hat{Z}$) for the three nested BraTS regions (WT, TC, ENH). Moreover, the synthesized volume is provided to a radiologist for blinded clinical validation, ensuring that the framework delivers both automated segmentation and verifiable visual context.}
 \label{fig:inference_pipeline}
\end{figure}

Addressing these limitations, we propose CAPS, a novel Clinically-Verifiable Anatomy-Prioritized Synthesis framework that bridges the gap between computational robustness and clinical reliability. Instead of optimizing solely for pixel-level similarity, our approach explicitly maximizes the preservation of anatomical structures and tumor topology. By restoring the missing modality with high structural fidelity, our method provides the clinician with the necessary visual data to plan surgeries and verify AI-generated segmentations, while simultaneously supporting robust downstream automated segmentation, as illustrated in Figure \ref{fig:inference_pipeline}.

The main contributions of this paper are summarized as follows:

\begin{itemize}
\item \textbf{Empirical Analysis of the Accuracy-Interpretability Trade-off.} We compare robust segmentation frameworks against synthesis-based pipelines and conclude that, while robust segmenters may achieve marginally higher numerical accuracy on missing modalities, high-fidelity image synthesis is an indispensable step to provide the visual context that neuroradiologists require for diagnostic verification.
\item \textbf{Anatomy-Prioritized Synthesis.} Unlike traditional approaches that rely solely on pixel-wise optimization, we design a generative architecture driven by a hybrid loss function ($\mathcal{L}_1+\mathcal{L}_{AFP}+\mathcal{L}_{DualDice}$) that explicitly enforces the preservation of fine anatomical structures, tumor boundaries, and overall integrity. Building upon the AFP loss originally proposed for MR-to-CT translation \cite{longuefosse2025anatomical}, we adapt it to the multi-modal MRI synthesis setting and complement it with a Dual Dice loss that jointly penalizes region-level and boundary-level deviations. Central to our design is a frozen, modality-specialized segmentation network that serves a dual supervisory role: as a multi-scale anatomical feature extractor and as a topological boundary enforcer, directly coupling the generative objective with downstream diagnostic consistency.
\item \textbf{Blinded Radiological Evaluation.} Moving beyond standard pixel-level metrics (e.g., PSNR, NMSE) that may not fully reflect clinical utility, we incorporate a blinded clinical validation. We conclude that our CAPS approach yields synthetic images with improved diagnostic quality and structural reliability compared to the standard synthesis baseline. To facilitate reproducibility and encourage clinical validation in future synthesis research, we publicly release the evaluation tool used for this assessment.
\item \textbf{Cross-Domain Generalization.} We successfully evaluate the synthesis capability of our method on completely unseen Meningioma, Pediatric Glioma, and Sub-Saharan Glioma datasets, effectively overcoming the domain shift without requiring any domain-specific adaptation.
\end{itemize}

\section{Related Work}

Handling missing mp-MRI modalities in neuro-oncology generally falls into two distinct paradigms: robust segmentation architectures that adapt to incomplete inputs \cite{zhang2024unified}, and medical image synthesis models that generate the absent data \cite{dar2019image}.

\subsection{Robust Segmentation with Missing Modalities}

Recent literature has extensively explored adaptive segmentation networks designed to be resilient against incomplete input configurations. Several studies have focused on learning modality-agnostic networks and shared latent representations that remain stable even when specific sequences are omitted \cite{dorent2019hetero}. For instance, early foundational approaches like HeMIS \cite{havaei2016hemis} achieved this by computing statistical moments across available feature maps. This concept has evolved into sophisticated feature-level fusion strategies employing attention mechanisms, transformer-based architectures like mmFormer \cite{zhang2022mmformer}, and state-space models such as IM-Fuse \cite{pipoli2025fuse}. Additionally, knowledge distillation techniques have been widely proposed to transfer structural information from a full-modality ``teacher'' network to a missing-modality ``student'' network \cite{hu2020knowledge}.

Despite achieving remarkably high computational accuracy, typically evaluated through standard metrics like the Dice Similarity Coefficient (DSC) and Hausdorff Distance (HD95), these robust segmenters face a critical usability limitation. Because they do not generate the missing scan and directly predict tissue boundaries, they fail to provide the structural reference that radiologists rely on. This lack of diagnostic transparency hinders the adoption of pure segmentation approaches in clinical workflows where visual verification is mandatory.

\subsection{Medical Image Synthesis}

A parallel line of research focuses on cross-modality medical image synthesis to impute the missing sequences. Early generative methods heavily relied on Generative Adversarial Networks (GANs) \cite{goodfellow2014generative, tavse2022systematic, al2024multi}, such as Pix2Pix \cite{isola2017image} and CycleGAN \cite{zhu2017unpaired}, which primarily performed one-to-one image translations. More recently, the field has advanced towards highly adaptable architectures capable of handling multi-modal inputs, shifting from basic 1-to-1 mappings to complex many-to-one synthesis models \cite{meng2024multi}. This evolution is highlighted by advanced Diffusion Models \cite{kebaili2025multi} like AMM-Diff \cite{kebaili2025amm}, which adaptively process variable sets of inputs, and State Space Models like I2I-Mamba \cite{atli2024i2i}, which capture long-range contextual features across multi-contrast MRI.

However, a fundamental limitation persists: most of these models are trained using generic pixel intensity-based objectives like $L_1$ or Mean Squared Error \cite{nie2018medical}. Because these losses penalize errors on a purely pixel basis, they lack structural awareness and tend to over-smooth the generated images, washing out critical high-frequency details such as intricate tumor borders \cite{isola2017image}. Furthermore, their performance is often solely measured via Image Quality Metrics (e.g., PSNR, SSIM), leaving their actual utility for downstream automated clinical tasks unverified.

\subsection{Structural Constraints and Joint Architectures}

To mitigate the smoothing effects of intensity-based losses, recent studies have introduced structural constraints and joint synthesis-segmentation frameworks. For example, methods like EaGAN \cite{yu2019ea} explicitly integrate edge-aware perceptual constraints to preserve high-frequency structures. Similarly, 3D Wavelet Diffusion Models (WDM) \cite{ferreira2024brain} strive to maintain spatial resolution while generating missing modalities. Moreover, unified frameworks have been proposed to jointly optimize synthesis and segmentation, acknowledging that image generation must ultimately be useful for tumor delineation \cite{dorent2025unified}. However, while feature-based objectives like perceptual loss effectively balance pixel accuracy and realism in natural scenes, they often lack the anatomical rigor required for 3D medical volumes, where structural integrity must be prioritized over stylistic consistency \cite{longuefosse2025anatomical}.

Despite these significant advances, existing joint frameworks often struggle to preserve fine tissue interfaces when critical modalities are missing. Our work addresses this exact issue. By integrating a targeted Anatomical Feature-Prioritized (AFP) loss and a Dual Dice loss into a multi-modal 3-to-1 generative architecture, our framework successfully reconciles algorithmic precision with clinical interpretability. This approach successfully delivers highly verifiable restorations for radiologists while maintaining competitive performance in subsequent segmentation tasks.

\section{Method}

This section details the proposed Clinically-Verifiable Anatomy-Prioritized Synthesis framework (CAPS), designed to impute missing MRI modalities while preserving high-fidelity structural details. First, the mathematical problem formulation and the overall two-stage pipeline are established. Subsequently, the network architecture and the hybrid objective function are introduced. Finally, the specific implementation and training configurations required to ensure robust convergence and downstream clinical utility are outlined.

\subsection{Problem Formulation and Pipeline}

The core objective of our framework is to synthesize a missing MRI scan from a subset of available complementary modalities, ensuring that the generated image retains both structural fidelity and diagnostic utility. 

Let $\mathcal{M} = \{T1n, T1c, T2w, T2f\}$ denote the complete set of four standard neuro-oncological MRI scan modalities acquired for a given patient. In real-world clinical scenarios, it is common for a specific modality to be missing, corrupted, or otherwise unavailable \cite{li2024brain}. We denote the target missing modality as $t \in \mathcal{M}$ and its corresponding ground-truth 3D volume as $Y$, and formalize the problem as a multi-modal synthesis task (specifically, a 3-to-1 mapping). The set of available input modalities is therefore defined as $\mathcal{X} = \mathcal{M} \setminus \{t\}$. Our goal is to train a generative mapping function $G$, parametrized by weights $\theta$, that takes the available modalities $\mathcal{X}$ as a multi-channel input and synthesizes a high-fidelity target estimation of the missing sequence, denoted as $\hat{Y}$:

\begin{equation}
    \hat{Y} = G_{\theta}(\mathcal{X}),
\end{equation}

\noindent where the synthesized volume $\hat{Y}$ is explicitly optimized to approximate the underlying anatomical information on $Y$.

\begin{figure*}[ht!]
\centering
\includegraphics[width=1.0\textwidth, trim=0cm 1.5cm 0cm 3cm, clip]{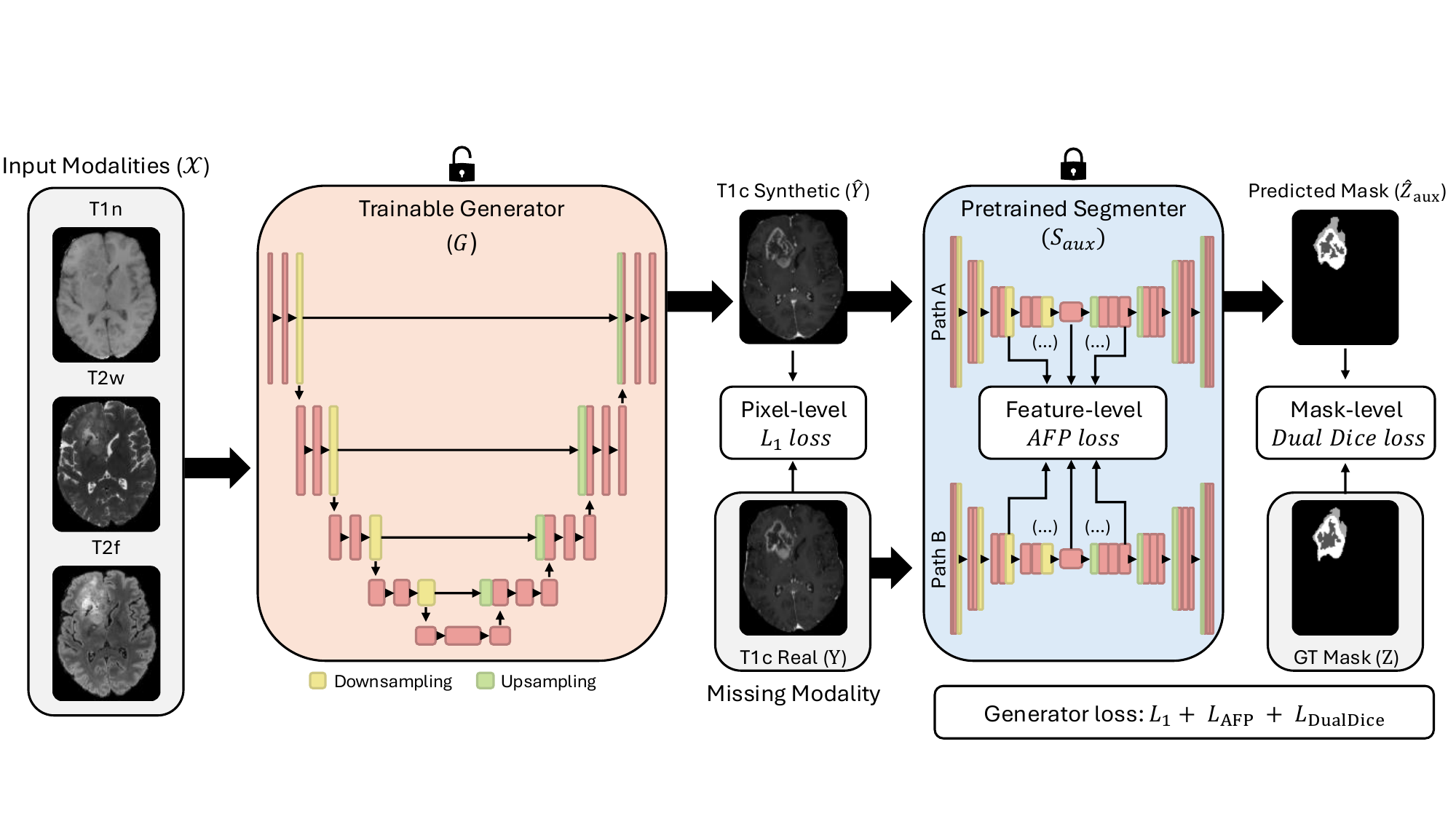}
\caption{\textbf{Overview of the Synthesis Framework.} A trainable generator $G$ synthesizes the missing modality ($\hat{Y}$, T1c in this example) from the available ones ($\mathcal{X}$: T1n, T2w, T2f) through a hybrid objective function. First, a pixel-level $\mathcal{L}_1$ loss enforces global intensity consistency between the synthesized volume $\hat{Y}$ and the real one $Y$. Then, both are processed by a frozen pre-trained auxiliary segmenter ($S_{aux}$), which acts as an expert feature extractor to compute the Anatomical Feature-Prioritized ($\mathcal{L}_{AFP}$) loss across multi-scale latent representations. Finally, the predicted mask ($\hat{Z}_{aux}$) and the real one ($Z$) are used to calculate the topological Dual Dice loss, thereby ensuring the preservation of both deep structural invariants and precise tumor boundaries.}
 \label{fig:AFP-training-strategy}
\end{figure*}

To achieve this, we propose a Clinically-Verifiable Anatomy-Prioritized Synthesis framework (CAPS) designed to operate in two interconnected phases (see Figure \ref{fig:inference_pipeline}):

\begin{enumerate}
    \item \textbf{Image Synthesis Phase:} The core generative network takes the available 3D volumes $\mathcal{X}$ and synthesizes $\hat{Y}$ according to Figure \ref{fig:AFP-training-strategy} scheme. Unlike traditional synthesis pipelines that optimize purely for intensity matching, our generator $G$ is driven by a hybrid objective function. This function integrates a pre-trained feature extractor to explicitly align the latent anatomical representations of $\hat{Y}$ with those of $Y$ through $S_{aux}$, ensuring the preservation of intricate tumor morphologies. 
    \item \textbf{Downstream Segmentation Phase:} To validate the clinical utility of the synthesized modality, $\hat{Y}$ is subsequently combined with the available modalities $\mathcal{X}$ to assemble the full imputed protocol, defined as $\mathcal{M}_{synth} = \mathcal{X} \cup \{\hat{Y}\}$. This complete set is then fed into an independent segmentation network $S$ to predict the final tumor boundaries. 
\end{enumerate}

This two-stage pipeline guarantees that the synthesis process is not evaluated in isolation, but rather through its ability to provide actionable and interpretable visual data that supports robust downstream clinical tasks.

It is important to note that, since each MRI sequence encodes fundamentally different tissue contrasts, both the generator $G$ and the auxiliary extractor $S_{aux}$ are independently trained for each specific missing modality scenario $t \in \mathcal{M}$, yielding four specialized pairs ($G_t$, $S_{aux,t}$). In contrast, the final clinical segmenter $S$ is trained as a single unified model capable of processing any imputed 4-modality protocol, regardless of which sequence was synthesized. This design ensures that each synthesis pathway is maximally specialized for its target modality, while the downstream segmenter remains a general-purpose clinical tool.

\subsection{CAPS Overview}

The proposed method is organized into three distinct components designed to prioritize anatomical fidelity: a generative trainable network $G$, an auxiliary feature extractor $\mathcal{S}_{aux}$, and a final clinical segmenter $S$. This tripartite design ensures that the synthesis process is guided by deep anatomical priors while remaining robust for downstream diagnostic tasks.

\subsubsection{Generator ($G$)}
The generator consists of a 3D Residual U-Net (nnResU-Net) \cite{isensee2024nnu} architecture configured with six encoding and decoding stages. 

By incorporating residual connections within each convolutional block, the network facilitates an efficient gradient flow that prevents the vanishing gradient problem typical of deep 3D models \cite{isensee2021nnu}, thereby accelerating convergence and enhancing the stability of the generative process. The encoder is designed to process a multi-channel stacked input $\mathcal{X}$, composed of the three available MRI modalities, while the decoder is responsible for synthesizing the single-channel volume $\hat{Y}$ corresponding to the missing sequence. To prevent the emergence of chequerboard artifacts, the decoder utilizes a trilinear interpolation strategy followed by standard convolutions \cite{longuefosse2026intensity}, ensuring smooth spatial transitions in the reconstructed anatomy.

\subsubsection{Auxiliary Feature Extractor ($\mathcal{S}_{aux}$)}
To drive the anatomy-prioritized optimization, we utilize an auxiliary network, $\mathcal{S}_{aux}$, which acts as a frozen expert during the training of $G$. This network is pre-trained using only one single real modality as input to ensure that its latent embeddings are highly specialized in the specific anatomical features of each sequence. During the generative phase, $\mathcal{S}_{aux}$ remains strictly frozen and serves as a multi-scale feature extractor. Specifically, to compute the Anatomical Feature-Prioritized ($\mathcal{L}_{AFP}$) loss \cite{longuefosse2025anatomical}, feature maps are extracted from every level of both the encoder and the decoder, forcing the generator to preserve structural invariants across multiple levels of abstraction. Furthermore, this network provides the topological guidance required for the $\mathcal{L}_{DualDice}$ loss, aligning the synthesized lesion area with the expected clinical morphology.

Architecturally, $\mathcal{S}_{aux}$ shares the same nnU-Net ResEnc M configuration and training objective as the final clinical segmenter $S$, with the critical distinction that its input layer is configured to accept a single modality channel rather than the full four-channel protocol.

\subsubsection{Final Clinical Segmenter ($S$)}
The final component of our framework is the clinical segmenter $S$, which utilizes the nnU-Net ResEnc M architecture \cite{isensee2024nnu} to achieve an optimal balance between segmentation performance and computational efficiency. This network is optimized using the standard nnU-Net objective function, combining Dice loss and Cross-Entropy loss in equal proportion. Unlike the auxiliary extractor, this network is trained to process the full complement of four input modalities. To enhance its resilience to synthetic imputation, $S$ is trained under a specialized augmentation scheme where there is a $50\%$ probability that any given modality is replaced by its synthetic counterpart generated by $G$. During the inference phase, this network acts as the definitive evaluator, processing the imputed 4-modality protocol to produce the final tumor masks. This configuration ensures that the segmenter is not only capable of interpreting synthetic data but also of integrating it seamlessly with real modalities to maintain high diagnostic accuracy.

\subsection{Loss Formulation}

The total objective function used to optimize the generator $G$ is formulated as a weighted sum of three distinct components: 1) a global intensity loss ($\mathcal{L}_{1}$); 2) an anatomical feature-level loss ($\mathcal{L}_{AFP}$); and 3) a topological segmentation loss ($\mathcal{L}_{DualDice}$). 
\begin{equation}
    \mathcal{L}_{Total} = \mathcal{L}_{1} + \mathcal{L}_{AFP} +  \mathcal{L}_{DualDice},
    \label{eq:generator_loss}
\end{equation}

\subsubsection{Voxel-Level Reconstruction Loss ($\mathcal{L}_{1}$)}
To ensure global intensity consistency and minimize the fundamental voxel-wise error between the synthesized modality $\hat{Y}$ and the ground truth one $Y$, we employ the standard $\mathcal{L}_1$ distance. Unlike Mean Squared Error ($L_2$), the $L_1$ norm is less prone to excessive blurring, making it a suitable baseline for contrast restoration \cite{isola2017image}. It is computed as:
\begin{equation}
    \mathcal{L}_{1} = \frac{1}{V} \sum_{v=1}^{V} \left| \hat{Y}_v - Y_v \right|,
\end{equation}
\noindent where $V$ is the total number of voxels in the 3D volume, and $\hat{Y}_v$ and $Y_v$ represent the intensity values of the $v$-th voxel in the predicted and target sequences, respectively.

\subsubsection{Anatomical Feature-Prioritized Loss 
($\mathcal{L}_{AFP}$)}
While $\mathcal{L}_1$ ensures global contrast, it lacks the spatial context required to preserve high-frequency structures. To explicitly promote the preservation of fine anatomical details, we employ the Anatomical Feature-Prioritized (AFP) loss, originally proposed for cross-modality translation in \cite{longuefosse2025anatomical} and here adapted to the multi-modal MRI synthesis setting. Instead of computing differences purely in the image space, $\mathcal{L}_{AFP}$ calculates the $L_1$ distance in a highly specialized latent space. 

By leveraging the pre-trained and frozen segmentation network $\mathcal{S}_{aux}$ as a feature extractor, we map both $\hat{Y}$ and $Y$ into their respective deep embeddings. The loss is defined as:
\begin{equation}
    \mathcal{L}_{AFP} = \frac{1}{N} \sum_{l=1}^{N} \left\| \phi_l(\hat{Y}) - \phi_l(Y) \right\|_1,
\end{equation}
\noindent where $N$ denotes the total number of extracted layers from $\mathcal{S}_{aux}$, and $\phi_l(\cdot)$ represents the intermediate feature activation maps obtained at the $l$-th layer.

\subsubsection{Dual Dice Loss ($\mathcal{L}_{DualDice}$)}

To enforce strict anatomical alignment specifically at the tumor regions, we propose a Dual Dice loss that evaluates both full-region overlap and boundary precision. 

Since $G$ outputs continuous intensity images, we map these volumes into a topological space by passing them through the frozen auxiliary segmentation network $\mathcal{S}_{aux}$. Let $\hat{Z}_{aux} = \mathcal{S}_{aux}(\hat{Y})$ represent the tumor mask predicted from the synthesized image, the loss is then computed directly against the manual clinical ground truth mask, denoted as $Z$. This formulation forces the generator to synthesize textures and intensities that are interpreted by the segmentation model in the same manner as the true anatomical pathology, thereby maximizing downstream diagnostic consistency.

To capture fine-grained morphological details, we define the boundary contours $\partial\hat{Z}_{aux}$ and $\partial Z$ through a differentiable morphological gradient $\Delta_k$. For any binary mask $B$, the boundary is extracted as $\partial B = \Delta_k(B) = \delta_k(B) - \epsilon_k(B)$, where $\delta_k$ and $\epsilon_k$ denote dilation and erosion operators with default kernel size $k=3$, implemented via 3D max-pooling. The proposed topological penalty is then formulated as:

\begin{equation}
    \mathcal{L}_{DualDice} = \left( 1 - \frac{2 |\hat{Z}_{aux} \cap Z|}{|\hat{Z}_{aux}| + |Z|} \right) + \left( 1 - \frac{2 |\partial\hat{Z}_{aux} \cap \partial Z|}{|\partial\hat{Z}_{aux}| + |\partial Z|} \right),
\end{equation}

\noindent where the first term ensures global topological agreement of the lesion, and the second term heavily penalizes morphological deviations strictly at the borders. This dual formulation prevents the generated tumor from artificially expanding or shrinking relative to the actual anatomy.

\subsection{Implementation and Training Details}

CAPS is implemented in PyTorch and integrated into the adapted nnU-Netv2 architecture \cite{isensee2021nnu, longuefosse2024adapted}. 

\paragraph{Preprocessing and Data Augmentation}
All MRI volumes were independently normalized using standard Z-score normalization (zero mean, unit variance) per modality. During training, we employed a robust online data augmentation pipeline to prevent overfitting, which included spatial transformations (random rotations and scaling) and intensity transformations (Gaussian noise, Gaussian blur, brightness/contrast adjustments, and gamma correction) \cite{isensee2021nnu}. Notably, spatial mirroring (flips) was explicitly disabled to preserve the inherent left-right anatomical asymmetries of the brain.

\paragraph{Training Configuration}
The generative model was trained using 3D patches of size $96 \times 160 \times 160$ voxels and a batch size of 2. The network parameters were optimized using Stochastic Gradient Descent (SGD) with Nesterov momentum \cite{shi2024hyperparameters} ($0.99$) and weight decay ($3 \times 10^{-5}$). The initial learning rate was set to $10^{-2}$ and progressively decayed according to a polynomial schedule function over the training process. 

To ensure training stability, especially given the multi-objective nature of our hybrid loss function, we applied gradient clipping. The model was trained for 1000 epochs, with each epoch consisting of 250 iterations. All models were trained on a single NVIDIA A40 GPU (48 GB VRAM).

\paragraph{Inference Phase}
During the evaluation phase, the inference was performed using a sliding-window approach with a $50\%$ patch overlap \cite{isensee2021nnu}. To reconstruct the final synthetic modality $\hat{Y}$, we employ a mean aggregation strategy across overlapping patches.

\section{Experimental Setup}

This section outlines the experimental configuration utilized to validate CAPS. Initially, the selected neuro-oncological datasets and their respective data partitioning strategies are detailed for both the primary evaluation and the cross-domain inference. Finally, the quantitative evaluation metrics employed to assess both image reconstruction quality and downstream segmentation accuracy are defined.

\subsection{Datasets and Data Partitioning}
To comprehensively train and evaluate our proposal, we utilized the globally recognized Brain Tumor Segmentation (BraTS) datasets \cite{menze2014multimodal}. All multi-parametric MRI sequences (T1n, T1c, T2w, and T2f) across these datasets share a standardized preprocessing pipeline: they are co-registered to the same anatomical template, skull-stripped, and resampled to an isotropic resolution of $1 \times 1 \times 1$ mm.

\paragraph{Primary Training and Evaluation (Adult Glioma)}
The core development of our method was conducted using the BraTS 2023 Adult Glioma dataset\cite{baid2021rsna} (conceptually identical to the BraTS 2021 and 2022 cohorts), whose official training set comprises $1251$ patient cases. We performed an $80\%-20\%$ split on this subset: the $80\%$ partition was exclusively used to train the generative synthesis network $G$, while the remaining $20\%$ was strictly reserved for training both the auxiliary feature extractor $S_{aux}$ and the final clinical segmenter $S$, ensuring complete data separation from the generator. 

In order to ensure a completely unbiased evaluation, all our quantitative results, ablation studies, and State-of-the-Art comparisons were extracted directly from the official, unseen BraTS Adult Glioma Validation set, which comprises unseen $219$ patient cases. To guarantee absolute fairness and reproducibility, the downstream segmentation metrics on this validation set were computed and retrieved through the official Synapse automated evaluation platform.

\paragraph{Cross-Domain Generalization Datasets}
To evaluate the zero-shot robustness of our framework under significant domain shifts, we conducted additional inference on three distinct, unseen neuro-oncological cohorts from the BraTS challenge:
\begin{itemize}
    \item BraTS Meningioma \cite{labella2023asnr}: $1000$ cases featuring structurally distinct, typically benign extra-axial tumors.
    \item BraTS Pediatric Glioma \cite{kazerooni2024brain}: $99$ cases, presenting unique developmental brain anatomies and differing tumor morphologies.
    \item BraTS Sub-Saharan Africa (SSA) Glioma \cite{adewole2023brain}: $60$ cases acquired from diverse, low-resource clinical settings, introducing significant variations in scanner hardware and image quality.
\end{itemize}

\subsection{Evaluation Metrics}

\paragraph{Image Synthesis Metrics}
To assess the quality of the imputed MRI modalities, we utilized three standard Image Quality Metrics (IQM): Normalized Mean Squared Error (NMSE), Peak Signal-to-Noise Ratio (PSNR), and Structural Similarity Index Measure (SSIM). 

Given the arbitrary and highly variable intensity ranges inherent to MRI sequences, computing these metrics directly on raw voxel values can yield inconsistent results. Therefore, prior to evaluation, both the predicted volume $\hat{Y}$ and the target volume $Y$ were robustly normalized to a standard $[0, 1]$ range using the 1st and 99th percentiles of the target volume intensity distribution  \cite{isensee2021nnu}. This step effectively clips hyper-intense artifacts and establishes a reliable dynamic range. Using the normalized volumes, the metrics are computed as follows:
\begin{equation}
    NMSE(Y, \hat{Y}) = \frac{\| Y - \hat{Y} \|_2^2}{\| Y \|_2^2}
\end{equation}
\begin{equation}
PSNR(Y, \hat{Y}) = 10 \cdot \log_{10} \left( \frac{MAX_Y^2}{MSE(Y, \hat{Y})} \right)
\end{equation}
\begin{equation}
    SSIM(Y, \hat{Y}) = \frac{(2\mu_Y \mu_{\hat{Y}} + c_1)(2\sigma_{Y\hat{Y}} + c_2)}{(\mu_Y^2 + \mu_{\hat{Y}}^2 + c_1)(\sigma_Y^2 + \sigma_{\hat{Y}}^2 + c_2)}
\end{equation}
\noindent where $\| \cdot \|_2^2$ denotes the squared $L_2$ norm, $MAX_Y = 1$ is the maximum possible pixel value after normalization, $\mu$ and $\sigma$ denote the mean and standard deviation of the images, and $c_1, c_2$ are standard stabilization constants. NMSE provides a robust measure of absolute intensity deviation, PSNR evaluates the global reconstruction quality relative to the peak signal, and SSIM assesses perceived textural and structural realism.

\paragraph{Segmentation Metrics}
To validate the clinical interpretability and functional utility of the synthesized images, we evaluated the performance of the segmentation networks using two complementary metrics. First, spatial overlap is measured by the standard Dice Similarity Coefficient (DSC):
\begin{equation}
    DSC(\hat{Z}, Z) = \frac{2 |\hat{Z} \cap Z|}{|\hat{Z}| + |Z|}
\end{equation}
\noindent where $\hat{Z}$ and $Z$ represent the predicted and ground truth multi-class tumor masks, respectively.

Second, to rigorously assess boundary precision, we employed the 95th percentile Hausdorff Distance (HD95). Unlike the maximum Hausdorff Distance, which is highly sensitive to single-voxel outliers \cite{taha2015metrics}, HD95 measures the spatial distance between the boundaries of the predicted mask ($\partial \hat{Z}$) and those of the true mask ($\partial Z$), disregarding the worst $5\%$ of matching errors:
\begin{equation}
    HD95(\partial \hat{Z}, \partial Z) = \max \left( d_{95}(\partial \hat{Z}, \partial Z), d_{95}(\partial Z, \partial \hat{Z}) \right)
\end{equation}
\noindent where $d_{95}(A, B)$ represents the 95th percentile of the minimum Euclidean distances from all points in contour $A$ to contour $B$.

\section{Results}

Evaluation of the proposed framework encompasses a rigorous analysis of both synthesis fidelity and clinical utility. Initial experiments establish the performance degradation caused by missing modalities, providing a baseline for the subsequent ablation studies on architecture and loss formulation. Additionally, the results include a comprehensive comparison with state-of-the-art models, alongside a zero-shot generalization analysis and a blinded radiological validation to confirm the diagnostic reliability of the synthesized modality.

\begin{table*}[!b]
\centering
\caption{\textbf{Quantitative Comparison of Baseline Strategies for Missing Modalities.} Segmentation performance (Dice and HD95) evaluated across different missing MRI scenarios. The \textit{Reference} model establishes the upper performance bound, having been trained and evaluated on the complete 4-modality protocol. For the missing scenarios, both \textit{Zero Imputation} and \textit{Synthesis Baseline} evaluate this exact same reference model by replacing the absent modality with zeros or a standard $\mathcal{L}_1$-synthesized volume, respectively. In contrast, \textit{Trained with Dropout} evaluates an independent baseline model specifically trained with modality dropout augmentation to handle incomplete inputs. The best performing baseline strategy for each specific missing scenario is highlighted in bold.}
\resizebox{0.8\textwidth}{!}{
\begin{tabular}{l cccc cccc}
\toprule
& \multicolumn{4}{c}{Dice ($\uparrow$)} & \multicolumn{4}{c}{HD95 ($\downarrow$)} \\
\cmidrule(lr){2-5} \cmidrule(lr){6-9}
Strategy & ENH & TC & WT & Mean & ENH & TC & WT & Mean \\
\midrule

\textit{Reference (All Real)} 
& \textit{81.47} & \textit{86.27} & \textit{92.22} & \textit{86.65} 
& \textit{21.56} & \textit{9.67} & \textit{4.98} & \textit{12.07} \\
\midrule
\midrule

\textbf{Missing T1n} & & & & & & & & \\
Zero Imputation 
& 79.60 & 83.55 & 91.59 & 84.91 & \textbf{22.37} & 12.84 & 5.71 & \textbf{13.64} \\
Trained with Dropout 
& 79.04 & \textbf{86.77} & \textbf{92.40} & \textbf{86.07} & 30.42 & \textbf{8.31} & \textbf{4.46} & 14.40 \\
Synthesis Baseline 
& \textbf{79.89} & 84.97 & 90.87 & 85.24 & 24.48 & 12.60 & 8.06 & 15.05 \\

\midrule

\textbf{Missing T1c} & & & & & & & & \\
Zero Imputation 
& 14.02 & 43.09 & 89.46 & 48.86 & 70.49 & 31.66 & 6.61 & 36.25 \\
Trained with Dropout 
& \textbf{58.81} & \textbf{77.38} & \textbf{91.95} & \textbf{76.05} & \textbf{35.59} & \textbf{12.56} & \textbf{4.60} & \textbf{17.58} \\
Synthesis Baseline 
& 52.27 & 70.27 & 90.03 & 70.86 & 41.44 & 18.83 & 7.37 & 22.55 \\

\midrule

\textbf{Missing T2w} & & & & & & & & \\
Zero Imputation 
& 77.26 & 79.81 & 80.45 & 79.17 & \textbf{27.77} & 18.26 & 16.34 & 20.29 \\
Trained with Dropout 
& \textbf{79.89} & \textbf{87.07} & \textbf{92.26} & \textbf{86.41} & 28.46 & \textbf{6.43} & \textbf{4.56} & \textbf{13.15} \\
Synthesis Baseline 
& 78.64 & 84.93 & 90.68 & 84.75 & 32.56 & 15.33 & 14.29 & 20.73 \\

\midrule

\textbf{Missing T2f} & & & & & & & & \\
Zero Imputation 
& 52.48 & 58.60 & 47.66 & 52.91 & 55.99 & 44.27 & 45.31 & 48.52 \\
Trained with Dropout 
& \textbf{79.42} & \textbf{86.47} & \textbf{89.91} & \textbf{85.27} & \textbf{28.58} & \textbf{8.08} & \textbf{5.76} & \textbf{14.14} \\
Synthesis Baseline 
& 75.13 & 83.28 & 89.09 & 82.50 & 36.44 & 13.68 & 8.61 & 19.58 \\

\midrule
\midrule

\textbf{Average (1 MRI missing)} & & & & & & & & \\
Zero Imputation 
& 55.84 & 66.26 & 69.03 & 63.71 & 44.16 & 26.76 & 18.49 & 29.80 \\
Trained with Dropout 
& \textbf{74.29} & \textbf{84.42} & \textbf{91.63} & \textbf{83.45} & \textbf{30.76} & \textbf{8.85} & \textbf{4.85} & \textbf{14.82} \\
Synthesis Baseline 
& 71.48 & 80.86 & 90.17 & 80.84 & 33.73 & 15.11 & 9.58 & 19.48 \\

\bottomrule
\end{tabular}
}
\label{tab:baseline_strategies}
\end{table*}

\subsection{Impact of Missing Modalities}

Before evaluating the results of CAPS, it is crucial to quantify the clinical and computational impact that the absence of MRI modalities has on automated segmentation systems, detailed in Table~\ref{tab:baseline_strategies}.

The reference model, trained and evaluated on the full protocol of four real modalities, establishes the upper performance bound with a mean Dice coefficient of $86.65\%$ and a Hausdorff distance (HD95) of $12.07$mm. On the other hand, the zero-imputation strategy reveals critical network dependencies on specific sequences, identifying the absence of the T1c sequence as the most critical scenario, where the Dice coefficient for the enhancing tumor (ENH) drastically collapses to $14.02\%$. Similarly, the missing T2f case disproportionately affects the whole tumor (WT), with its Dice score dropping to $47.66\%$ due to the loss of contrast in the peritumoral edema regions.

\begin{table*}[ht!]
\centering
\caption{\textbf{Generator Architecture ($G$) Ablation Study.} Quantitative performance comparison of different generative network architectures for the synthesis task in the missing T1c scenario. The study evaluates the impact of progressively upgrading the baseline nnU-Net by integrating 3D residual blocks (nnResU-Net) and replacing standard transposed convolutions with a trilinear interpolation decoder. Performance is assessed using both downstream clinical segmentation metrics (Dice and HD95) and standard Image Quality Metrics, IQM (NMSE and PSNR). The best results among the evaluated generative architectures are highlighted in bold.}
\resizebox{0.9\textwidth}{!}{
\begin{tabular}{l cccc cccc cc}
\toprule
& \multicolumn{4}{c}{Dice ($\uparrow$)} & \multicolumn{4}{c}{HD95 ($\downarrow$)} & \multicolumn{2}{c}{IQM} \\ 
\cmidrule(lr){2-5} \cmidrule(lr){6-9} \cmidrule(lr){10-11} 
Network & ENH & TC & WT & Mean & ENH & TC & WT & Mean & NMSE ($\downarrow$) & PSNR ($\uparrow$) \\ 
\midrule

{\textit{Reference (All Real)}} 
& \textit{81.47} & \textit{86.27} & \textit{92.22} & \textit{86.65}
& \textit{21.56} & \textit{9.67} & \textit{4.98} & \textit{12.07}
& - & - \\ 
\midrule

nnU-net & 52.27 & 70.27 & 90.03 & 70.86 & 41.44 & 18.83 & \textbf{7.37} & 22.55 & 0.0261 & 27.60 \\ 
nnResU-net  & 54.47 & \textbf{71.50} & 90.01 & 71.99 & \textbf{38.44} & 19.22 & 7.47 & \textbf{21.71}  & 0.0256 & 27.72  \\ 
nnResUnet + trilinear decoder & \textbf{54.54} & 71.46 & \textbf{90.26} & \textbf{72.09} & 41.13 & \textbf{17.42} & 7.47 & 22.01  & \textbf{0.0241} & \textbf{28.05} \\ 
\bottomrule
\end{tabular}}
\label{tab:best-trainer}
\end{table*}

\begin{table*}[!b]
\centering
\caption{\textbf{Objective Loss Function Ablation Study.} Quantitative evaluation of different loss configurations for the optimized generative network $G$ (nnResU-Net with trilinear decoder) in the missing T1c scenario. The baseline pixel-wise $\mathcal{L}_1$ loss is progressively augmented with the Anatomical Feature-Prioritized ($\mathcal{L}_{AFP}$) loss and topological Dice constraints ($\text{Dice}_1$ targeting global region overlap, and $\text{Dice}_2$ focusing on boundary precision). The proposed hybrid objective ($\mathcal{L}_1 + \mathcal{L}_{AFP} + \text{Dice}_{1,2}$), obtained with the combined use of Generator $G$ and pretrained segmenter $S_{aux}$, achieves the best overall downstream segmentation performance, obtaining the highest mean Dice score and lowest mean HD95.}
\resizebox{0.9\textwidth}{!}{
\begin{tabular}{l c ccc cccc cc} 
\toprule
& \multicolumn{4}{c}{Dice ($\uparrow$)} & \multicolumn{4}{c}{HD95 ($\downarrow$)} & \multicolumn{2}{c}{IQM} \\ 
\cmidrule(lr){2-5} \cmidrule(lr){6-9} \cmidrule(lr){10-11} 
Loss & ENH & TC & WT & Mean & ENH & TC & WT & Mean & NMSE ($\downarrow$) & PSNR ($\uparrow$) \\ 
\midrule

{\textit{Reference (All Real)}}
& \textit{81.47} & \textit{86.27} & \textit{92.22} & \textit{86.65}
& \textit{21.56} & \textit{9.67} & \textit{4.98} & \textit{12.07}
& - & - \\ 
\midrule

L1 
 & 54.54 & 71.46 & 90.26 & 72.09 & 41.13 & 17.42 & 7.47 & 22.01  & 0.0241 & \textbf{28.05}\\ 

L1 + $Dice_1$   
& 54.83 & 73.70 & 90.97 & 73.17 & 39.74 & 15.32 & 7.01 & 20.69 & 0.0286 & 26.95  \\ 

L1 + $Dice_2$    
& 54.93 & \textbf{76.99} & 91.17 & 74.36 & 38.21 & 13.07 & \textbf{5.94} & 19.07 & 0.0290 & 26.88  \\ 

L1 + AFP   
& 55.33 & 73.46 & \textbf{91.46} & 74.42 & 37.59 & 14.53 & 6.26 & 19.46 & \textbf{0.0240} & 27.83  \\ 

L1 + AFP + $Dice_1$   & 56.10 & 75.41 & 90.09 & 73.87 & 38.56 & 15.66 & 7.47 & 20.56 & 0.0245 & 27.63 \\

L1 + AFP + $Dice_2$ 
& 54.80 & 76.21 & 90.50 & 73.84 & 45.14 & \textbf{12.17} & 6.64 & 21.32 & 0.0254 & 27.49  \\ 

L1 + AFP + $Dice_{1,2}$   
& \textbf{57.19} & 76.18 & 90.10 & \textbf{74.49} & \textbf{37.22} & 12.45 & 7.08 & \textbf{18.92} & 0.0259 & 27.32  \\

\bottomrule
\end{tabular}
}
\label{tab:loss-ablation}
\end{table*}

Although robust approaches, such as modality dropout training, partially mitigate these declines, these systems operate under a ``black-box'' paradigm that delivers an isolated segmentation mask without restoring the underlying anatomical volume. This lack of visual evidence prevents neuroradiologists from corroborating the AI-generated segmentation, thus justifying the pressing need for our synthesis framework, which is aimed at preserving structural fidelity alongside diagnostic interpretability.

\subsection{Ablation Study: Architecture and Loss}

To empirically justify the design choices of our synthesis and segmentation framework, we conducted a comprehensive ablation study where all configurations were evaluated under the missing T1c scenario, as it represents the most critical challenge from an anatomical perspective, according to Table~\ref{tab:baseline_strategies}. The architectural and loss decisions derived from this scenario were subsequently applied identically to the remaining three missing modality cases, with final results reported across all scenarios in Table \ref{tab:results}.

\paragraph{Architectural Optimization of $G$}

As detailed in Table~\ref{tab:best-trainer}, we established a baseline using the standard nnU-Net architecture, which yielded a mean Dice coefficient of $70.86\%$ and a HD95 of $22.55$ mm. The integration of 3D residual blocks (nnResU-Net) facilitated a more robust training optimization, raising the Dice score to $71.99\%$. This performance gain is characteristic of residual architectures, which help mitigate gradient degradation in deep 3D networks while preserving high-frequency spatial features. Finally, replacing standard transposed convolutions with a trilinear interpolation decoder effectively mitigated spatial chequerboard artifacts, stabilizing synthesis quality and achieving an optimal NMSE, PSNR and mean dice.

\paragraph{Hybrid Loss Formulation through $S_{aux}$}

\begin{figure*}[t]
\centering
\includegraphics[width=1.0\textwidth, trim=0cm 0cm 0cm 0cm, clip]{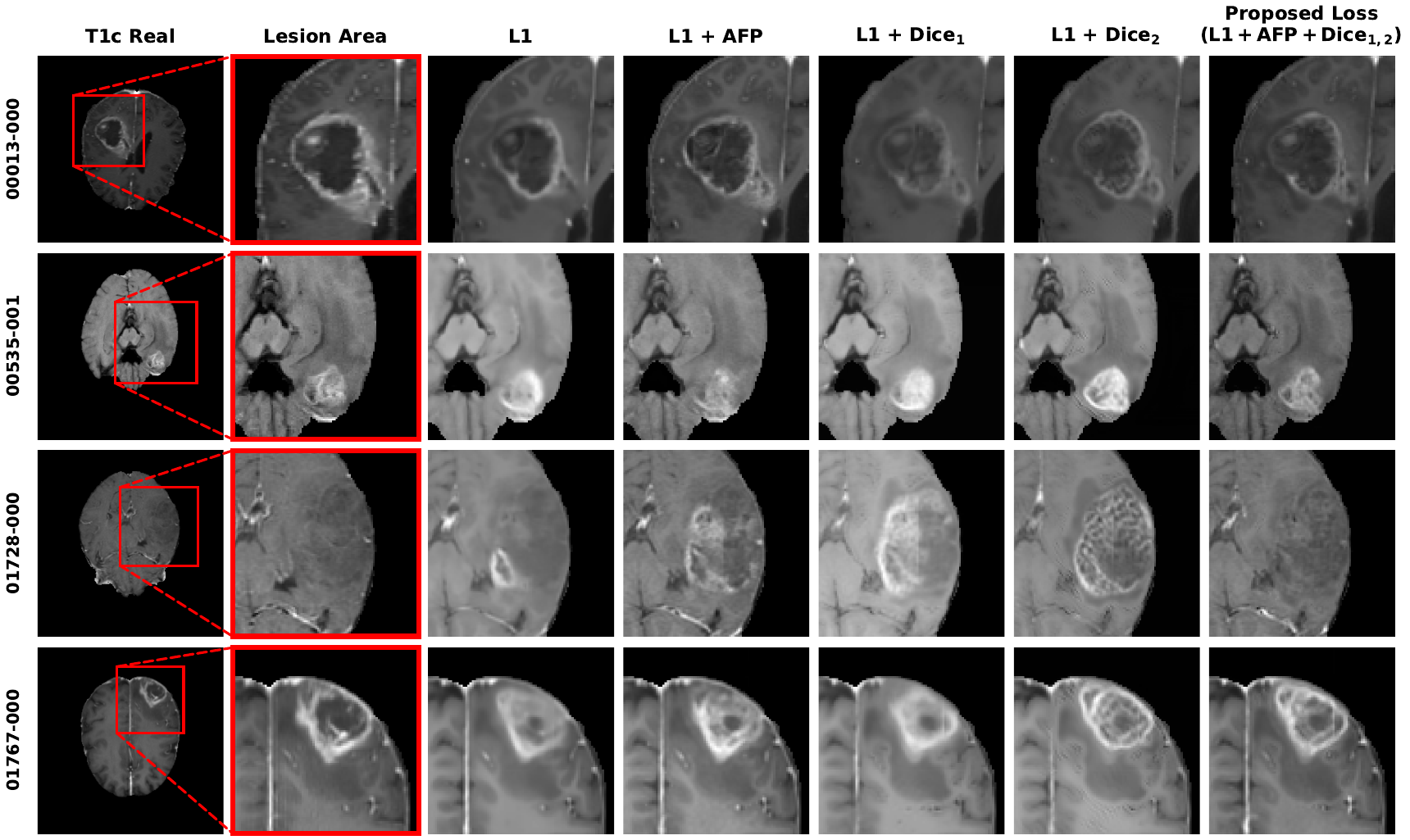}
\caption{\textbf{Visual Comparison of Loss Configurations.} Generated T1c sequences under varying loss configurations for the missing T1c scenario. The purely pixel-wise $\mathcal{L}_1$ optimization produces over-smoothed lesions lacking structural definition. While intermediate configurations ($\mathcal{L}_1 + \mathcal{L}_{AFP}$ and variations with partial Dice constraints) yield incremental improvements in boundary demarcation, the proposed hybrid loss ($\mathcal{L}_1 + \mathcal{L}_{AFP} + \mathcal{L}_{DualDice}$) achieves the optimal qualitative balance. It successfully recovers both the sharp anatomical margins and the critical intra-tumoral heterogeneity and the regions of neoangiogenesis, closely mirroring the real T1c target.}
 \label{fig:loss_generator_visual}
\end{figure*}

The quantitative results presented in Table~\ref{tab:loss-ablation} and the visual comparisons in Figure~\ref{fig:loss_generator_visual} underscore the critical importance of our hybrid loss formulation. As observed in Table~\ref{tab:loss-ablation}, optimization relying exclusively on the voxel-level absolute error ($\mathcal{L}_1$) yields the poorest segmentation performance, which suggests that purely intensity-based metrics do not prioritize the preservation of essential anatomical structures, treating all voxels with equal weight regardless of their diagnostic relevance.

The introduction of Dice-based components into the objective function leads to a measurable improvement in the mean Dice score of the synthesized images. However, the most robust validation results are achieved only when combining all three terms: $\mathcal{L}_1$, the Anatomical Feature-Prioritized loss ($\mathcal{L}_{AFP}$), and the $\mathcal{L}_{DualDice}$, which ensures a synergy between global intensity consistency, deep latent feature alignment, and topological boundary integrity.

Qualitatively, Figure~\ref{fig:loss_generator_visual} demonstrates that $\mathcal{L}_1$-driven reconstructions suffer from significant blurring, particularly at the tumor margins. While intermediate solutions, such as  $\mathcal{L}_1 + AFP$, succeed in producing well-defined boundaries that closely resemble the ground truth, our proposed total loss configuration ($\mathcal{L}_1 + \mathcal{L}_{AFP} + \mathcal{L}_{DualDice}$) consistently provides the superior balance. Specifically for T1c sequences, where the identification of active tumor zones and neoangiogenesis is clinically vital, our framework successfully preserves these intricate anatomical details while simultaneously generating sharp, non-blurry volumes that mirror the target distribution.

\paragraph{Segmentation Strategy on $S$}

Having optimized the image generator $G$, we evaluated the most effective strategy for integrating these synthetic volumes into the segmentation network $S$. We compared training under varying missing modality ratios (25\%, 50\%, 75\%) using Random Dropout, where the T1c modality is zero-padded with the given probability, against a generative augmentation approach, where the missing T1c is replaced by our model's synthetic imputation. All configurations were ultimately evaluated in the scenario where the 3 real MRI modalities are available and the T1c modality must be synthesized.

The underlying hypothesis of the Random Dropout strategy is that, by forcing the network to encounter zero-padded T1c channels during training, it will develop fewer dependencies on this specific modality, theoretically improving robustness when it is missing at inference. However, our results demonstrate that this approach is suboptimal, likely due to the inherent domain shift between the training conditions (zero-padded inputs) and the testing conditions (synthesized inputs). 

Instead, we observed that explicitly feeding the segmentation network with synthetically imputed volumes during training consistently yields superior performance. Specifically, training the segmenter from scratch with a 50\% probability of replacing the real T1c with its synthetic counterpart during training provided the best overall results, particularly in improving the mean Dice coefficient (75.80\%). 

\begin{table*}[ht!]
\centering
\caption{\textbf{Segmenter ($S$) Training Strategy Ablation.} Quantitative evaluation of different training paradigms for the final clinical segmenter under the missing T1c scenario. Using the optimized generator configuration, we explore the impact of varying the probability ratio (25\%, 50\%, and 75\%) of encountering a missing modality during training. The standard \textit{Random Dropout} (zero-padding) baseline is compared against our proposed synthetic data augmentation, evaluated both by training the segmenter from scratch and by fine-tuning (FT) a pre-trained model. The results demonstrate that training the segmenter from scratch with a 50\% probability of synthetic imputation achieves the optimal balance, yielding the highest mean Dice score and validating the utility of generative augmentation for downstream robustness.}
\resizebox{0.8\textwidth}{!}{
\begin{tabular}{l cccc cccc}
\toprule
& \multicolumn{4}{c}{Dice ($\uparrow$)} & \multicolumn{4}{c}{HD95 ($\downarrow$)} \\
\cmidrule(lr){2-5} \cmidrule(lr){6-9}
Training Strategy & ENH & TC & WT & Mean & ENH & TC & WT & Mean \\
\midrule

{\textit{Reference (All Real)}}
& \textit{81.47} & \textit{86.27} & \textit{92.22} & \textit{86.65}
& \textit{21.56} & \textit{9.67} & \textit{4.98} & \textit{12.07} \\

\midrule
\midrule
\multicolumn{9}{l}{\textbf{25\% Missing Modality Ratio}} \\

Random Dropout 
& 57.39 & 77.10 & 90.89 & 75.13 & 40.19 & 13.47 & 6.24 & 19.97 \\
Synthetic (from scratch)
& 56.98 & 77.55 & \textbf{92.05} & 75.53 & \textbf{35.98} & \textbf{10.85} & \textbf{4.44} & \textbf{17.09} \\
Synthetic (FT) 
& 56.76 & 76.40 & 90.13 & 74.43 & 37.56 & 14.89 & 11.44 & 21.30 \\

\midrule

\multicolumn{9}{l}{\textbf{50\% Missing Modality Ratio}} \\

Random Dropout 
& 56.73 & 76.86 & 90.93 & 74.84 & 38.69 & 13.58 & 7.52 & 19.93 \\
Synthetic (from scratch)
& \textbf{57.95} & \textbf{77.59} & 91.86 & \textbf{75.80} & 36.10 & 10.90 & 4.63 & 17.21  \\
Synthetic (FT)
& 55.10 & 75.96 & 90.33 & 73.80 & 39.88 & 15.88 & 8.70 & 21.49 \\

\midrule

\multicolumn{9}{l}{\textbf{75\% Missing Modality Ratio}} \\

Random Dropout 
& 57.67 & 77.18 & 90.36 & 75.07 & 38.81 & 14.92 & 11.68 & 21.80 \\
Synthetic (from scratch) 
& 58.25 & 76.88 & 90.98 & 75.37 & 37.09 & 12.18 & 5.48 & 18.25   \\
Synthetic (FT)
& 55.59 & 75.90 & 90.39 & 73.96 & 40.47 & 14.43 & 7.67 & 20.86 \\

\bottomrule
\end{tabular}
}
\label{tab:segmenter-ablation}
\end{table*}

\paragraph{Overall Performance Across All Modalities}
Having established the optimal generator architecture $G$, hybrid loss formulation through $S_{aux}$, and downstream segmentation strategy on $S$ for the missing T1c scenario, we extended the evaluation of our final framework to all other missing modality cases, presented in Table \ref{tab:results}.

The results confirm that the benefits of our methodology are not isolated to T1c, but generalize exceptionally well to any missing sequence. For instance, in the missing T1n and T2w scenarios, the framework achieves mean Dice scores of $85.90\%$ and $85.12\%$, respectively, alongside high PSNR values ($28.99$ and $26.18$). Even in the missing T2f scenario, which heavily impacts edema delineation, the model successfully restores the peritumoral contrast, yielding a mean Dice of $82.69\%$. Overall, our unified framework achieves a highly robust global average Dice of $82.38\%$ and an HD95 of $15.96$ mm across all missing modality scenarios, solidifying its reliability as a comprehensive imputation tool for incomplete clinical protocols.

\begin{table*}[!ht]
\centering
\caption{\textbf{Overall Performance Across All Missing Modalities.} Evaluation of the fully optimized CAPS for every single missing MRI scenario (T1n, T1c, T2w, and T2f). The table reports both clinical segmentation accuracy (Dice and HD95) and Image Quality Metrics, IQM (NMSE and PSNR). Utilizing the optimal generative architecture $G$, hybrid loss formulation through the use of $S_{aux}$, and synthetic data augmentation strategy on $S$ established in the previous ablation studies, the framework demonstrates robust imputation fidelity across all sequences, achieving a global average Dice score of $82.38\%$.}
\resizebox{0.8\textwidth}{!}{
\begin{tabular}{
l
@{\hskip 10pt} c @{\hskip 8pt} c @{\hskip 8pt} c @{\hskip 8pt} c
@{\hskip 10pt} c @{\hskip 8pt} c @{\hskip 8pt} c @{\hskip 8pt} c
@{\hskip 10pt} c @{\hskip 8pt} c
}
\toprule
& \multicolumn{4}{c}{Dice ($\uparrow$)}
& \multicolumn{4}{c}{HD95 ($\downarrow$)}
& \multicolumn{2}{c}{IQM}\\
\cmidrule(lr){2-5} \cmidrule(lr){6-9} \cmidrule(lr){10-11}
& ENH & TC & WT & Mean & ENH & TC & WT & Mean & NMSE ($\downarrow$) & PSNR ($\uparrow$) \\
\midrule

Missing T1n
& 80.58 & 85.87 & 91.24 & 85.90
& 22.56 & 11.07 & 6.02 & 13.22
& 0.0160 & 28.99 \\

Missing T1c
& 57.95 & 77.59 & 91.86 & 75.80 & 36.10 & 10.90 & 4.63 & 17.21  & 0.0259 & 27.32  \\

Missing T2w
& 79.22 & 85.46 & 90.69 & 85.12
& 22.99 & 11.59 & 10.91 & 15.16
& 0.0530 & 26.18 \\

Missing T2f
& 73.58 & 83.93 & 90.57 & 82.69
& 36.93 & 10.09 & 7.68 & 18.23
& 0.0453 & 25.34 \\

\midrule
\textit{Average}
& \textit{72.83} & \textit{83.21} & \textit{91.09} & \textit{82.38}
& \textit{29.65} & \textit{10.91} & \textit{7.31} & \textit{15.96}
& \textit{0.0351} & \textit{26.96} \\

\bottomrule
\end{tabular}
}
\label{tab:results}
\end{table*}

\subsection{Comparison with State-of-the-Art}
To comprehensively validate our Synthesis Framework, we compared its performance against recent State-of-the-Art approaches in both image reconstruction quality and downstream segmentation utility.

\paragraph{Image Reconstruction Performance}
Table~\ref{tab:sota_reconstruction} presents a quantitative comparison of synthesis quality. A persistent challenge in this domain is the lack of standardized evaluation protocols on the BraTS dataset; existing literature diverges significantly in terms of target missing modalities, data partitioning (e.g., using the official validation set versus reserving an internal 20\% held-out split from the training data), and input availability (1-to-1 vs. 3-to-1 synthesis) \cite{li2024brain}. Furthermore, the limited availability of public source code for several baselines precludes perfectly matched retraining. To ensure the most rigorous and fair evaluation possible under these constraints, we carefully aligned our testing with the specific conditions reported by each baseline.

In the official validation split, our volumetric approach significantly outperforms existing 1-to-1 methods (e.g., Pix2Pix \cite{isola2017image}, CycleGAN \cite{zhu2017unpaired}, EaGAN \cite{yu2019ea}, ALDM \cite{kim2024adaptive}), achieving an SSIM of $93.97\%$. While this performance gap naturally underscores the substantial advantage of utilizing the full multi-modal context (3-to-1 mapping) inherently supported by our architecture, it also highlights the capacity of our model to effectively integrate complementary information for high-fidelity imputation. To further evaluate our architectural robustness and isolate the effect of input configuration, we compared our method against recent 3-to-1 models, such as AMM-Diff \cite{kebaili2025amm}, under their reported evaluation protocol (a 20\% held-out split). Evaluated on this corresponding fraction of unseen data, our method maintains a highly competitive SSIM of $93.88\%$. Even though diffusion-based models like AMM-Diff \cite{kebaili2025amm} achieve a marginally higher PSNR, our approach excels in SSIM, demonstrating remarkable structural consistency.

\begin{table*}[ht!]
\centering
\caption{\textbf{State-of-the-Art Comparison of Image Reconstruction Quality.} Quantitative synthesis evaluation against existing baselines using standard Image Quality Metrics (IQM), reported as the mean across all evaluated missing modality scenarios within each group. To ensure a fair comparison, methods are grouped according to their specific evaluated missing scenarios and reported data partitions (Official Validation set vs. a 20\% held-out split). The results demonstrate that CAPS, our proposed 3-to-1 framework, substantially outperforms traditional 1-to-1 methods and achieves highly competitive structural fidelity (highest SSIM) compared to recent multi-modal diffusion approaches.}
\label{tab:sota_reconstruction}
\small
\resizebox{0.8\textwidth}{!}{
\begin{tabular}{l l c c c c c}
\toprule
 & & \multicolumn{3}{c}{IQM} & \multicolumn{2}{c}{Synthesis Type} \\
\cmidrule(lr){3-5} \cmidrule(lr){6-7}
Method & Eval. Split & NMSE $\downarrow$ & PSNR $\uparrow$ & SSIM \% $\uparrow$ & 1-to-1 & 3-to-1 \\
\midrule

\multicolumn{7}{l}{\textit{Evaluated Scenarios: Missing T2w, T2f}} \\
\midrule
Pix2Pix \cite{isola2017image} & Official Val. & 0.105 & 24.89 & 87.05 & $\times$ & \\
CycleGAN \cite{zhu2017unpaired} & Official Val. & 0.145 & 23.62 & 82.85 & $\times$ & \\
EaGAN \cite{yu2019ea} & Official Val. & 0.103 & 24.73 & 86.65 & $\times$ & \\
ALDM \cite{kim2024adaptive} & Official Val. & 0.089 & 25.45 & 88.55 & $\times$ & \\
CAPS & Official Val. & \textbf{0.049} & \textbf{25.76} & \textbf{93.97} & & $\times$ \\

\midrule
\multicolumn{7}{l}{\textit{Evaluated on All Missing Modality Scenarios (T1n, T1c, T2w, T2f)}} \\
\midrule
Pix2Pix \cite{isola2017image} & 20\% Held-out Split & & 29.70 & 75.83 & & $\times$ \\
AMM-Diff \cite{kebaili2025amm} & 20\% Held-out Split & & \textbf{31.80} & 93.50 & & $\times$ \\
CAPS & 20\% Held-out Split & \textbf{0.036} & 27.12 & \textbf{93.88} & & $\times$ \\

\bottomrule
\end{tabular}}
\end{table*}

\paragraph{Downstream Segmentation Performance}
Beyond pure synthesis quality, the ultimate goal of our framework is to provide data that is functionally useful for clinical workflows. Table~\ref{tab:sota_segmentation} details the segmentation accuracy of our synthesis-based pipeline against two distinct paradigms: robust segmentation frameworks and other synthesis-based approaches.

As expected, pure robust segmentation architectures (such as mmFormer \cite{zhang2022mmformer} and IM-Fuse \cite{pipoli2025fuse}) establish the upper numerical bounds, achieving mean Dice scores of $86.93\%$ and $87.34\%$, respectively. Because these models bypass the complex task of image generation and directly optimize boundary predictions from incomplete data, their purely computational metrics are higher. However, as previously discussed, this comes at the severe cost of clinical interpretability, outputting isolated masks without the underlying anatomical context.

Evaluating our framework within the interpretable synthesis-based paradigm presents additional challenges, as the majority of existing literature focuses exclusively on reconstruction metrics and does not report downstream segmentation performance. When compared to the available synthesis-based joint models that do provide these metrics, such as the recent work by Ferreira et al. \cite{ferreira2024brain}, our approach demonstrates a substantial improvement, raising the mean Dice score from $74.67\%$ to $82.38\%$. 

\begin{figure*}[!b]
\centering
\includegraphics[width=1.0\textwidth, trim=0cm 0cm 0cm 0cm, clip]{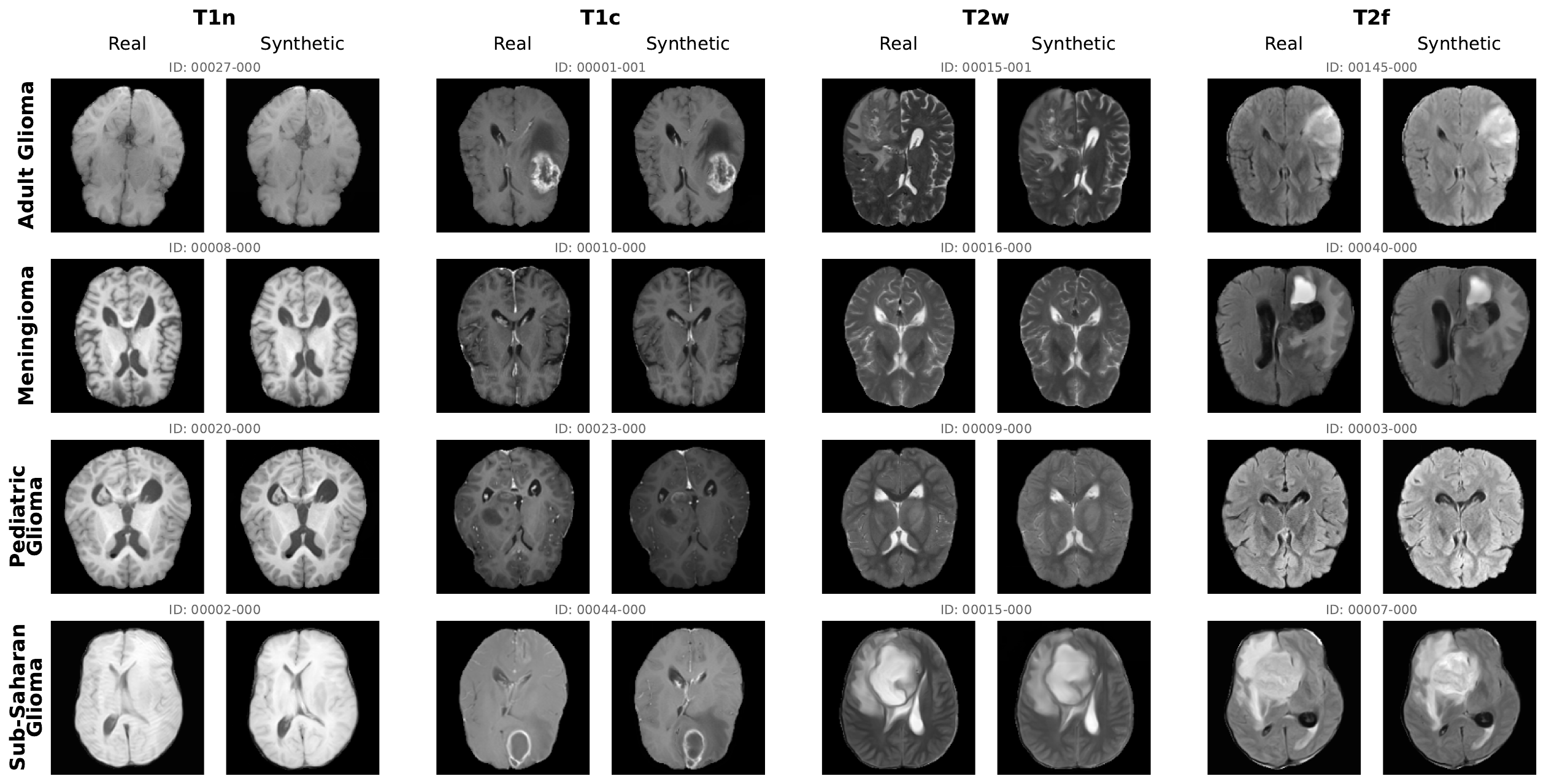}
\caption{\textbf{Zero-Shot Cross-Domain Generalization.} Qualitative comparison of real and synthesized MRI modalities (T1n, T1c, T2w, T2f) across the in-distribution training cohort (Adult Glioma) and three completely unseen Out-of-Distribution datasets: Meningioma, Pediatric Glioma, and Sub-Saharan African Glioma. CAPS consistently reconstructs the missing sequences with high structural fidelity across all scenarios, thereby reproducing domain-specific morphological variations and demonstrating robust generalization capabilities without requiring any domain-specific adaptation.}
\label{fig:visual_results}
\end{figure*}

Notably, while pure segmentation frameworks achieve approximately a $5\%$ higher mean Dice score, they operate as ``black boxes'', forcing clinicians to blindly trust the automated boundaries without the underlying visual context \cite{zhu2026no}. CAPS deliberately trades this purely computational advantage for a critical gain in clinical interpretability. By explicitly enforcing structural and topological consistency via our hybrid loss, we deliver highly verifiable anatomical restorations alongside robust tumor delineations. This approach ensures that neuroradiologists have the necessary visual evidence to validate the predictions of our method.

\begin{table}[ht!]
\centering
\caption{\textbf{State-of-the-Art Comparison of Segmentation.} Evaluation of tumor segmentation accuracy (Dice scores), reported as the average across all four missing modality scenarios (T1n, T1c, T2w, and T2f). Methods are categorized into pure robust segmentation models and interpretable 3-to-1 synthesis-based pipelines. Our framework outperforms existing synthesis-based approaches and remains competitive with leading purely computational segmenters.}
\label{tab:sota_segmentation}
\small
\resizebox{0.47\textwidth}{!}{
\begin{tabular}{ l c c c c }
\toprule
Method & ENH $\uparrow$ & TC $\uparrow$ & WT $\uparrow$ & Mean $\uparrow$  \\
\midrule
\textit{Robust Segmentation:} & & & & \\
Random Dropout & 74.29 & 84.42 & 91.63 & 83.45 \\
U-HVED \cite{dorent2019hetero} & 70.88 & 80.75 & 88.63 & 80.08 \\
mmFormer \cite{zhang2022mmformer} & 80.23 & 88.38 & 92.20 & 86.93 \\
Sfusion \cite{liu2023sfusion} & 77.95 & 86.60 & 91.05 & 85.20 \\
IM-Fuse \cite{pipoli2025fuse} & \textbf{80.73} & \textbf{88.95} & \textbf{92.35} & \textbf{87.34} \\
\midrule
\textit{Synthesis-based:} & & & &  \\
\cite{ferreira2024brain} & 68.00 & 73.00 & 83.00 & 74.67   \\
CAPS & \textbf{72.83} & \textbf{83.21} & \textbf{91.09} & \textbf{82.38}   \\
\bottomrule
\end{tabular}
}
\end{table}

\subsection{Cross-Domain Generalization}

\begin{table*}[ht!]
\centering
\caption{\textbf{Zero-Shot Cross-Domain Generalization.} Quantitative synthesis evaluation (NMSE and PSNR) on three completely unseen external datasets: Meningioma, Pediatric Glioma, and Sub-Saharan African Glioma. CAPS maintains consistently high image quality across all missing modality scenarios. This demonstrates robust out-of-the-box generalization to morphological and technical domain shifts, entirely without the need for domain-specific adaptation.}
\resizebox{0.8\textwidth}{!}{
\begin{tabular}{l cc cc cc}
\toprule
& \multicolumn{2}{c}{Meningioma} & \multicolumn{2}{c}{Pediatric Glioma} & \multicolumn{2}{c}{Sub-Saharan Glioma} \\
\cmidrule(lr){2-3} \cmidrule(lr){4-5} \cmidrule(lr){6-7}
Missing Modality & NMSE ($\downarrow$) & PSNR ($\uparrow$) & NMSE ($\downarrow$) & PSNR ($\uparrow$) & NMSE ($\downarrow$) & PSNR ($\uparrow$) \\
\midrule

4MRI Real & 0.0 & $\infty$ & 0.0 & $\infty$ & 0.0 & $\infty$ \\

\noalign{\vspace{3pt}} 
\hdashline
\noalign{\vspace{3pt}} 

Missing T1n & 0.0248 & 27.72 & 0.0216 & 28.05 & 0.0215 & 27.56 \\
Missing T1c & 0.0505 & 25.41 & 0.0330 & 25.93 & 0.0272 & 25.91 \\
Missing T2w & 0.0534 & 26.21 & 0.0648 & 25.20 & 0.0507 & 25.67 \\
Missing T2f & 0.0360 & 25.63 & 0.0266 & 26.46 & 0.0279 & 26.55 \\
\midrule
\textit{Average} & \textit{0.0412} & \textit{26.24} & \textit{0.0365} & \textit{26.41} & \textit{0.0318} & \textit{26.42} \\

\bottomrule
\end{tabular}
}
\label{tab:cross_domain_synthesis}
\end{table*}

\begin{figure*}[!b]
\centering
\includegraphics[width=1.0\textwidth, trim=0cm 0cm 0cm 0cm, clip]{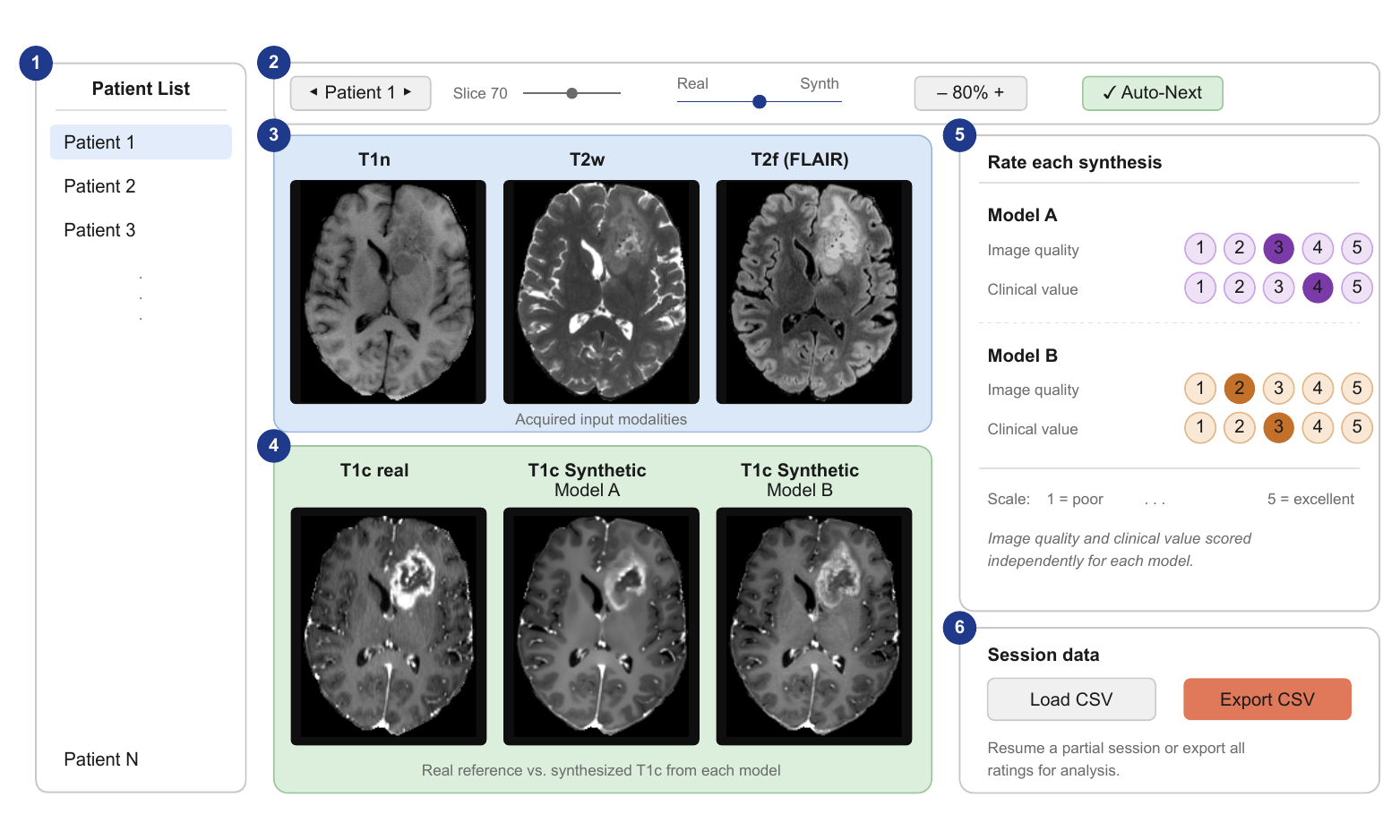}
\caption{\textbf{Blinded Radiological Evaluation Interface.} Web-based tool for clinical validation of synthesized T1c volumes, organized into six panels: (1) patient list for case navigation; (2) slice selector with Real/Synth overlay toggle and windowing controls; (3) available input modalities (T1n, T2w, T2f); (4) side-by-side comparison of the real T1c against two anonymized synthetic volumes (Model A and Model B), whose assignment is randomly shuffled per patient to prevent positional bias; (5) independent scoring of Image Quality and Clinical Value on a 1--5 scale for each model; and (6) session management for exporting and resuming evaluations. \\ Tool available at \url{https://github.com/Phyrise/NIfTI-to-HTML-Evaluator}.}
\label{fig:validation_tool}
\end{figure*}

A critical limitation of many medical deep learning frameworks is their susceptibility to domain shift. Models trained on a specific cohort often experience severe performance degradation when exposed to Out-of-Distribution (OoD) data \cite{diana2026gbt}. To validate the robustness of our framework, we evaluated its zero-shot inference capabilities on three completely unseen datasets from the BraTS challenge: Meningioma \cite{labella2023asnr}, Pediatric Glioma \cite{kazerooni2024brain}, and Sub-Saharan African Glioma \cite{adewole2023brain}.

Each of these datasets introduces unique morphological and technical challenges. Meningiomas are tumors that typically tend to compress rather than infiltrate surrounding brain tissue, presenting a fundamentally different topology than adult gliomas \cite{fathi2013meningioma}. Pediatric Gliomas involve developing brains with distinct myelination patterns and tumor presentations \cite{duffau2022white}. Finally, the Sub-Saharan African Glioma cohort introduces significant technical variability, featuring scans acquired from diverse hardware and protocols typical of low-resource clinical settings \cite{boonzaier2025domain}.

As reported in Table~\ref{tab:cross_domain_synthesis}, the framework demonstrates robust generalization despite these substantial domain shifts. Notably, the average synthesis quality on the three unseen cohorts (PSNR ranging from 26.24 to 26.42; NMSE from 0.0318 to 0.0412) remains closely aligned with the in-distribution Adult Glioma performance reported in Table \ref{tab:results} (PSNR = 26.96; NMSE = 0.0351). Furthermore, the narrow spread across the three external datasets indicates that the fidelity of our method is not contingent on a specific tumor morphology or acquisition protocol. Downstream segmentation metrics are intentionally omitted for these cross-domain cohorts for two reasons. First, the BraTS Meningioma dataset employs a fundamentally different tumor annotation protocol, rendering segmentation predictions from our Adult Glioma-trained segmenter $S$ inapplicable. Second, for the Pediatric and Sub-Saharan cohorts, reporting segmentation accuracy would conflate two independent sources of performance variation: synthesis fidelity and segmenter generalization, thereby preventing a clean assessment of the generator's cross-domain robustness, which is the specific focus of this evaluation.

Qualitatively, Figure~\ref{fig:visual_results} corroborates these numerical findings by presenting visual comparisons of real and synthesized volumes across all four modality scenarios (T1n, T1c, T2w, and T2f). The inference is shown for both the in-distribution Adult Glioma cohort and the three unseen OoD domains. As observed, the generated images closely mirror the real anatomical targets. Despite the unique anatomical and technical presentations of each dataset, the framework successfully preserves the overall brain structures and intricate tumor topologies without noticeable degradation.

\subsection{Clinical Validation}

\begin{figure}[ht!]
\centering
\includegraphics[width=0.45\textwidth, trim=0cm 0cm 0cm 0cm, clip]{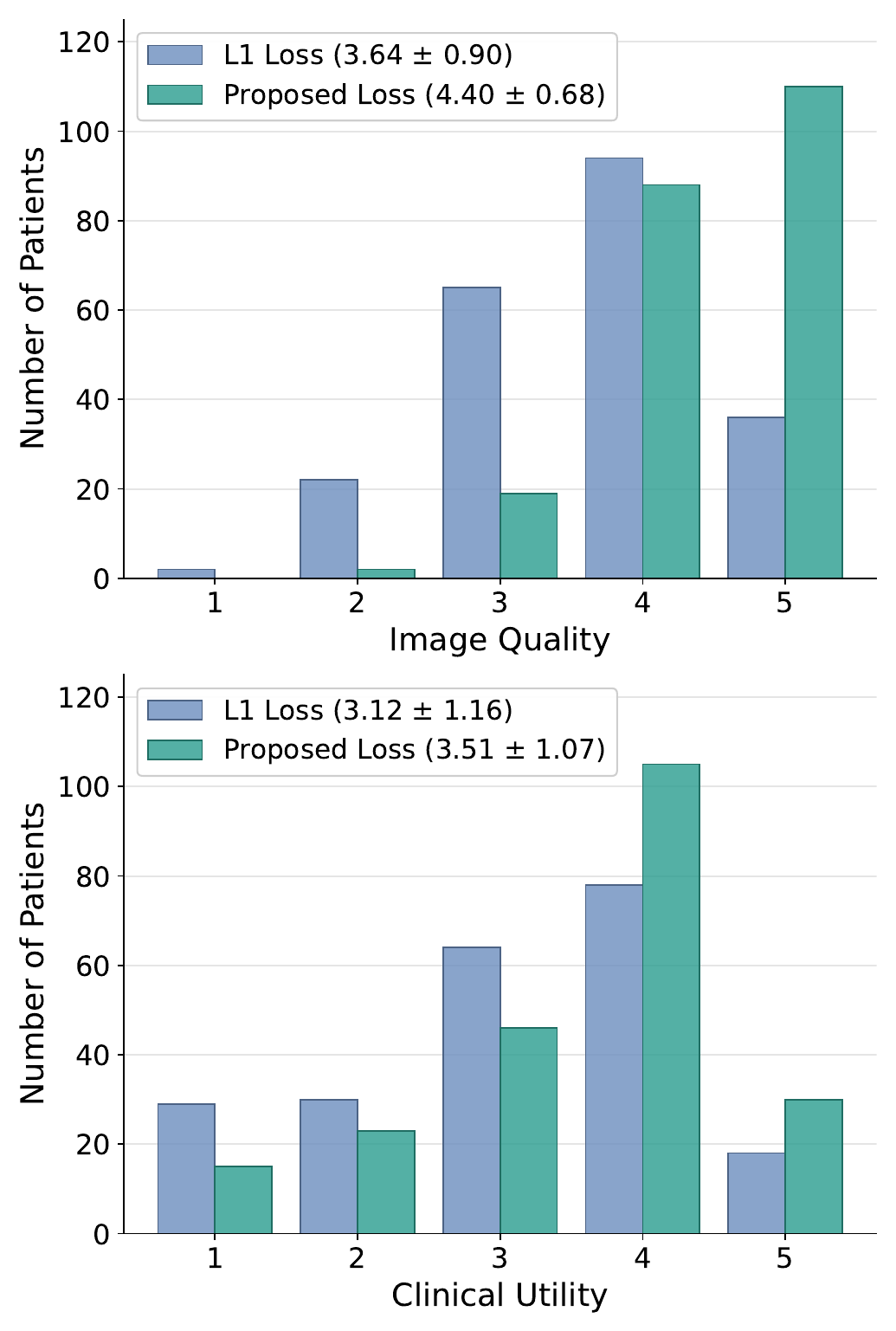}
\caption{\textbf{Radiologist Evaluation Scores.} Distribution of blinded radiologist scores comparing the standard $\mathcal{L}_1$ baseline (blue) against our proposed synthesis generation (green). To facilitate a direct visual comparison, the data is presented as follows: the upper half displays the score distribution for \textit{Image Quality}, while the lower half displays the scores for \textit{Clinical Utility}. As observed, the proposed hybrid loss decisively shifts the distribution towards the maximum scores (4-5) in both evaluation dimensions, demonstrating its superior visual fidelity and diagnostic reliability.}
 \label{fig:clinical_validation}
\end{figure}

While Image Quality Metrics (IQMs) are standard benchmarks in the  synthesis literature, they do not inherently guarantee that a generated 
image is diagnostically useful or structurally trustworthy for 
clinicians. To assess the practical clinical value of the synthesized 
images, we conducted a blinded radiological evaluation.

A senior radiology resident evaluated the generated T1c volumes for the 
complete Adult Glioma validation set (219 patients). The evaluator was 
presented with volumes synthesized using both the baseline pixel-wise 
objective ($\mathcal{L}_1$) and our proposed hybrid objective 
($\mathcal{L}_1 + \mathcal{L}_{AFP} + \mathcal{L}_{DualDice}$). The 
assessment was conducted in a blinded manner, without revealing which 
model generated which image, using the evaluation tool illustrated in 
Figure~\ref{fig:validation_tool}. For each patient, the assignment of 
synthesis methods to Model A and Model B was randomly shuffled to prevent positional bias. The radiologist scored each volume on a 1-5 scale across two distinct axes: \textit{Image Quality} (encompassing contrast, noise, and the presence of artifacts) and \textit{Clinical Utility}.

As illustrated in Figure~\ref{fig:clinical_validation}, the 
$\mathcal{L}_1$ baseline \textit{Image Quality} scores were concentrated 
between 3 and 4 ($\mu=3.64 \pm 0.90$), while the proposed hybrid loss 
shifted the distribution significantly toward scores of 4 and 5 
($\mu=4.40 \pm 0.68$). A consistent trend was observed for 
\textit{Clinical Utility}, where the proposed method achieved a score of 
$3.51 \pm 1.07$, outperforming the baseline's $3.12 \pm 1.16$.

To rigorously quantify the significance of these improvements, we 
conducted paired Wilcoxon signed-rank tests across all $219$ evaluated 
patients. The proposed hybrid loss yielded a statistically significant 
improvement over the $\mathcal{L}_1$ baseline in both evaluation axes: 
Image Quality ($W = 1080$, $p = 6.50 \times 10^{-24}$, $r = 0.68$) and 
Clinical Utility ($W = 1837$, $p = 7.93 \times 10^{-10}$, $r = 0.42$), 
with significance maintained after Bonferroni correction for multiple 
comparisons ($\alpha_{\text{adj}} = 0.025$). The effect sizes indicate a 
large improvement in perceived image quality and a medium improvement in 
clinical utility. Beyond the higher mean scores, the proposed loss also 
exhibits a lower standard deviation across both axes, reflecting enhanced robustness across diverse clinical presentations. These results confirm that our anatomy-aware synthesis provides reliable visual evidence to strengthen clinical confidence and support surgical planning.

\paragraph{Limitations $\&$ Future Work}
Despite these promising results, the current method presents certain limitations, particularly its decoupled two-stage structure and its reliance on a specific pre-trained segmentation network. To address these constraints, future research could focus on the development of a unified, end-to-end architecture that leverages anatomy-prioritized constraints within a single joint framework, potentially further closing the accuracy-interpretability gap. Additionally, other promising avenues for investigation include adapting the framework to accommodate dynamic input configurations (e.g., 1-to-1 or 2-to-1 synthesis) and extending this anatomy-prioritized paradigm to other complex oncological domains, such as multi-parametric prostate \cite{armato2018prostatex} or liver MRI. Finally, although the clinical validation confirms the diagnostic utility of the synthesized volumes, the assessment was performed by a single radiologist, so future studies may incorporate multiple experts to robustly assess inter-rater reliability. 

\section{Conclusion}

In this work, we presented a Clinically-Verifiable Anatomy-Prioritized Synthesis framework (CAPS) to address the critical challenge of missing MRI modalities in neuro-oncological imaging \cite{chrysochoou2026ai}. Traditional approaches to this problem typically force a compromise between computational accuracy and diagnostic transparency: robust segmenters predict tumor boundaries while leaving clinicians blind to the underlying anatomy \cite{zhang2022mmformer, pipoli2025fuse}, whereas standard synthesis models reconstruct images that lack the structural fidelity required for clinical trust \cite{longuefosse2025anatomical}. By driving our 3D generative architecture with a hybrid objective function ($\mathcal{L}_1 + \mathcal{L}_{AFP} + \mathcal{L}_{DualDice}$), we successfully bridged this gap, enforcing the preservation of deep anatomical features and strict topological boundaries. 

Our comprehensive evaluation on the BraTS Adult Glioma dataset \cite{baid2021rsna} demonstrated that our framework significantly outperforms existing State-of-the-Art synthesis methods with a highly competitive downstream segmentation mean Dice score of $82.38\%$. Beyond standard computational metrics, our approach exhibited exceptional zero-shot generalization capabilities across completely unseen and morphologically distinct domains, including Meningioma \cite{labella2023asnr}, Pediatric \cite{kazerooni2024brain}, and Sub-Saharan African \cite{adewole2023brain} cohorts. Most importantly, a blinded radiological evaluation confirmed that our anatomy-aware synthesis substantially improves both perceived image quality and clinical utility, providing human experts with the verifiable visual context needed to trust downstream automated segmentation.

\hspace{2pt}

\textbf{Funding Statements.} This work has been partially supported by the Agencia Estatal de Investigación of Spain under Grant PID2025-171806OB-I00 and by the Regional Government of Madrid of Spain under Grant TEC-2024/COM-322, and by the FPI-UAM grant.

\hspace{2pt}

\textbf{Data Availability}
The BraTS 2023 Adult Glioma, Meningioma, Pediatric Glioma, and Sub-Saharan African Glioma datasets used in this study are publicly available through the Synapse platform (\url{https://www.synapse.org/Synapse:syn51156910}). 

\hspace{2pt}

\textbf{Code Availability}
The generative framework was developed by adapting the publicly available nnU-Net translation codebase (\url{https://github.com/Phyrise/nnUNet_translation}), which provides the modified nnU-Net architecture for MR-to-CT synthesis including the AFP loss implementation. The final clinical segmenter $S$ was built upon the official nnU-Net repository (\url{https://github.com/MIC-DKFZ/nnUNet}). The blinded radiological evaluation tool is publicly available at \url{https://github.com/Phyrise/NIfTI-to-HTML-Evaluator}

\hspace{2pt}

\textbf{CRediT Authorship Contribution Statement}
Cecilia Diana-Albelda: Conceptualization, Methodology, Software, 
Validation, Formal Analysis, Investigation, Writing - Original Draft, Writing - Review \& Editing, Visualization. Arthur Longuefosse: Conceptualization, Methodology, Software, Validation, Investigation, Writing - Review \& Editing, Visualization. Mario Rub\'{e}n Paz Campos: Investigation, Validation (clinical radiological evaluation). \'{A}lvaro Garc\'{i}a-Mart\'{i}n: Conceptualization, Supervision, Writing - Review \& Editing, Funding Acquisition. Jesus Bescos: Conceptualization, Supervision, Writing - Review \& Editing,  Funding Acquisition.

\hspace{2pt}

\textbf{Declaration of Competing Interests}
The authors declare that they have no known competing financial 
interests or personal relationships that could have appeared to 
influence the work reported in this paper.








\bibliographystyle{elsarticle-harv}

\bibliography{references}

\begin{thebibliography}{47}
\expandafter\ifx\csname natexlab\endcsname\relax\def\natexlab#1{#1}\fi
\providecommand{\url}[1]{\texttt{#1}}
\providecommand{\href}[2]{#2}
\providecommand{\path}[1]{#1}
\providecommand{\DOIprefix}{doi:}
\providecommand{\ArXivprefix}{arXiv:}
\providecommand{\URLprefix}{URL: }
\providecommand{\Pubmedprefix}{pmid:}
\providecommand{\doi}[1]{\href{http://dx.doi.org/#1}{\path{#1}}}
\providecommand{\Pubmed}[1]{\href{pmid:#1}{\path{#1}}}
\providecommand{\bibinfo}[2]{#2}
\ifx\xfnm\relax \def\xfnm[#1]{\unskip,\space#1}\fi
\bibitem[{Adewole et~al.(2023)Adewole, Rudie, Gbdamosi, Toyobo, Raymond, Zhang, Omidiji, Akinola, Suwaid, Emegoakor et~al.}]{adewole2023brain}
\bibinfo{author}{Adewole, M.}, \bibinfo{author}{Rudie, J.D.}, \bibinfo{author}{Gbdamosi, A.}, \bibinfo{author}{Toyobo, O.}, \bibinfo{author}{Raymond, C.}, \bibinfo{author}{Zhang, D.}, \bibinfo{author}{Omidiji, O.}, \bibinfo{author}{Akinola, R.}, \bibinfo{author}{Suwaid, M.A.}, \bibinfo{author}{Emegoakor, A.}, et~al., \bibinfo{year}{2023}.
\newblock \bibinfo{title}{The brain tumor segmentation (brats) challenge 2023: Glioma segmentation in sub-saharan africa patient population (brats-africa)}.
\newblock \bibinfo{journal}{ArXiv} , \bibinfo{pages}{arXiv--2305}.
\bibitem[{Al~Khalil et~al.(2024)Al~Khalil, Ayaz, Lorenz, Weese, Pluim and Breeuwer}]{al2024multi}
\bibinfo{author}{Al~Khalil, Y.}, \bibinfo{author}{Ayaz, A.}, \bibinfo{author}{Lorenz, C.}, \bibinfo{author}{Weese, J.}, \bibinfo{author}{Pluim, J.}, \bibinfo{author}{Breeuwer, M.}, \bibinfo{year}{2024}.
\newblock \bibinfo{title}{Multi-modal brain tumor segmentation via conditional synthesis with fourier domain adaptation}.
\newblock \bibinfo{journal}{Computerized Medical Imaging and Graphics} \bibinfo{volume}{112}, \bibinfo{pages}{102332}.
\bibitem[{Armato~III et~al.(2018)Armato~III, Huisman, Drukker, Hadjiiski, Kirby, Petrick, Redmond, Giger, Cha, Mamonov et~al.}]{armato2018prostatex}
\bibinfo{author}{Armato~III, S.G.}, \bibinfo{author}{Huisman, H.}, \bibinfo{author}{Drukker, K.}, \bibinfo{author}{Hadjiiski, L.}, \bibinfo{author}{Kirby, J.S.}, \bibinfo{author}{Petrick, N.}, \bibinfo{author}{Redmond, G.}, \bibinfo{author}{Giger, M.L.}, \bibinfo{author}{Cha, K.}, \bibinfo{author}{Mamonov, A.}, et~al., \bibinfo{year}{2018}.
\newblock \bibinfo{title}{Prostatex challenges for computerized classification of prostate lesions from multiparametric magnetic resonance images}.
\newblock \bibinfo{journal}{Journal of Medical Imaging} \bibinfo{volume}{5}, \bibinfo{pages}{044501--044501}.
\bibitem[{Atli et~al.(2024)Atli, Kabas, Arslan, Demirtas, Yurt, Dalmaz and Cukur}]{atli2024i2i}
\bibinfo{author}{Atli, O.F.}, \bibinfo{author}{Kabas, B.}, \bibinfo{author}{Arslan, F.}, \bibinfo{author}{Demirtas, A.C.}, \bibinfo{author}{Yurt, M.}, \bibinfo{author}{Dalmaz, O.}, \bibinfo{author}{Cukur, T.}, \bibinfo{year}{2024}.
\newblock \bibinfo{title}{I2i-mamba: Multi-modal medical image synthesis via selective state space modeling}.
\newblock \bibinfo{journal}{arXiv preprint arXiv:2405.14022} .
\bibitem[{Baid et~al.(2021)Baid, Ghodasara, Mohan, Bilello, Calabrese, Colak, Farahani, Kalpathy-Cramer, Kitamura, Pati et~al.}]{baid2021rsna}
\bibinfo{author}{Baid, U.}, \bibinfo{author}{Ghodasara, S.}, \bibinfo{author}{Mohan, S.}, \bibinfo{author}{Bilello, M.}, \bibinfo{author}{Calabrese, E.}, \bibinfo{author}{Colak, E.}, \bibinfo{author}{Farahani, K.}, \bibinfo{author}{Kalpathy-Cramer, J.}, \bibinfo{author}{Kitamura, F.C.}, \bibinfo{author}{Pati, S.}, et~al., \bibinfo{year}{2021}.
\newblock \bibinfo{title}{The rsna-asnr-miccai brats 2021 benchmark on brain tumor segmentation and radiogenomic classification}.
\newblock \bibinfo{journal}{arXiv preprint arXiv:2107.02314} .
\bibitem[{Boonzaier et~al.(2025)Boonzaier, Moosa, Lebang, Jabaar, Iorumbur, Zhang and Raymond}]{boonzaier2025domain}
\bibinfo{author}{Boonzaier, W.P.}, \bibinfo{author}{Moosa, F.}, \bibinfo{author}{Lebang, K.}, \bibinfo{author}{Jabaar, H.}, \bibinfo{author}{Iorumbur, A.}, \bibinfo{author}{Zhang, D.}, \bibinfo{author}{Raymond, C.}, \bibinfo{year}{2025}.
\newblock \bibinfo{title}{Domain adaptation for adult glioma segmentation in sub-saharan africa: An ensemble of nnu-net v2 and mednext}, in: \bibinfo{booktitle}{International Conference on Medical Image Computing and Computer-Assisted Intervention}, \bibinfo{organization}{Springer}. pp. \bibinfo{pages}{248--261}.
\bibitem[{Chrysochoou et~al.(2026)Chrysochoou, Gandhi, Adib, Familiar, Vunnava, Varshochi, Khalili, Khalili, Ware, Tu et~al.}]{chrysochoou2026ai}
\bibinfo{author}{Chrysochoou, D.}, \bibinfo{author}{Gandhi, D.B.}, \bibinfo{author}{Adib, S.}, \bibinfo{author}{Familiar, A.M.}, \bibinfo{author}{Vunnava, B.}, \bibinfo{author}{Varshochi, S.}, \bibinfo{author}{Khalili, N.}, \bibinfo{author}{Khalili, N.}, \bibinfo{author}{Ware, J.B.}, \bibinfo{author}{Tu, W.}, et~al., \bibinfo{year}{2026}.
\newblock \bibinfo{title}{Ai-powered segmentation and prognosis with missing mri in pediatric brain tumors}.
\newblock \bibinfo{journal}{npj Precision Oncology} .
\bibitem[{Dar et~al.(2019)Dar, Yurt, Karacan, Erdem, Erdem and Cukur}]{dar2019image}
\bibinfo{author}{Dar, S.U.}, \bibinfo{author}{Yurt, M.}, \bibinfo{author}{Karacan, L.}, \bibinfo{author}{Erdem, A.}, \bibinfo{author}{Erdem, E.}, \bibinfo{author}{Cukur, T.}, \bibinfo{year}{2019}.
\newblock \bibinfo{title}{Image synthesis in multi-contrast mri with conditional generative adversarial networks}.
\newblock \bibinfo{journal}{IEEE transactions on medical imaging} \bibinfo{volume}{38}, \bibinfo{pages}{2375--2388}.
\bibitem[{Diana-Albelda et~al.(2026)Diana-Albelda, Alcover-Couso, Garc{\'\i}a-Mart{\'\i}n, Bescos and Escudero-Vi{\~n}olo}]{diana2026gbt}
\bibinfo{author}{Diana-Albelda, C.}, \bibinfo{author}{Alcover-Couso, R.}, \bibinfo{author}{Garc{\'\i}a-Mart{\'\i}n, {\'A}.}, \bibinfo{author}{Bescos, J.}, \bibinfo{author}{Escudero-Vi{\~n}olo, M.}, \bibinfo{year}{2026}.
\newblock \bibinfo{title}{Gbt-sam: A parameter-efficient depth-aware model for generalizable brain tumor segmentation on mp-mri}.
\newblock \bibinfo{journal}{International Journal of Imaging Systems and Technology} \bibinfo{volume}{36}, \bibinfo{pages}{e70389}.
\bibitem[{Diana-Albelda et~al.(2025a)Diana-Albelda, Garc{\'\i}a-Mart{\'\i}n and Bescos}]{diana2025review}
\bibinfo{author}{Diana-Albelda, C.}, \bibinfo{author}{Garc{\'\i}a-Mart{\'\i}n, {\'A}.}, \bibinfo{author}{Bescos, J.}, \bibinfo{year}{2025}a.
\newblock \bibinfo{title}{A review on deep learning methods for glioma segmentation, limitations, and future perspectives}.
\newblock \bibinfo{journal}{Journal of Imaging} \bibinfo{volume}{11}, \bibinfo{pages}{269}.
\bibitem[{Diana-Albelda et~al.(2025b)Diana-Albelda, Longuefosse, Garc{\'\i}a-Mart{\'\i}n and Bescos}]{diana2025multi}
\bibinfo{author}{Diana-Albelda, C.}, \bibinfo{author}{Longuefosse, A.}, \bibinfo{author}{Garc{\'\i}a-Mart{\'\i}n, {\'A}.}, \bibinfo{author}{Bescos, J.}, \bibinfo{year}{2025}b.
\newblock \bibinfo{title}{Multi-modal brain mri synthesis with nnu-net: Exploring segmentation performance and cross-modality relationships}, in: \bibinfo{booktitle}{International Workshop on Simulation and Synthesis in Medical Imaging}, \bibinfo{organization}{Springer}. pp. \bibinfo{pages}{52--62}.
\bibitem[{Dorent et~al.(2025)Dorent, Haouchine, Golby, Frisken, Kapur and Wells}]{dorent2025unified}
\bibinfo{author}{Dorent, R.}, \bibinfo{author}{Haouchine, N.}, \bibinfo{author}{Golby, A.}, \bibinfo{author}{Frisken, S.}, \bibinfo{author}{Kapur, T.}, \bibinfo{author}{Wells, W.}, \bibinfo{year}{2025}.
\newblock \bibinfo{title}{Unified cross-modal medical image synthesis with hierarchical mixture of product-of-experts}.
\newblock \bibinfo{journal}{IEEE Transactions on Pattern Analysis and Machine Intelligence} .
\bibitem[{Dorent et~al.(2019)Dorent, Joutard, Modat, Ourselin and Vercauteren}]{dorent2019hetero}
\bibinfo{author}{Dorent, R.}, \bibinfo{author}{Joutard, S.}, \bibinfo{author}{Modat, M.}, \bibinfo{author}{Ourselin, S.}, \bibinfo{author}{Vercauteren, T.}, \bibinfo{year}{2019}.
\newblock \bibinfo{title}{Hetero-modal variational encoder-decoder for joint modality completion and segmentation}, in: \bibinfo{booktitle}{International Conference on Medical Image Computing and Computer-Assisted Intervention}, \bibinfo{organization}{Springer}. pp. \bibinfo{pages}{74--82}.
\bibitem[{Duffau(2022)}]{duffau2022white}
\bibinfo{author}{Duffau, H.}, \bibinfo{year}{2022}.
\newblock \bibinfo{title}{White matter tracts and diffuse lower-grade gliomas: the pivotal role of myelin plasticity in the tumor pathogenesis, infiltration patterns, functional consequences and therapeutic management}.
\newblock \bibinfo{journal}{Frontiers in Oncology} \bibinfo{volume}{12}, \bibinfo{pages}{855587}.
\bibitem[{Fathi and Roelcke(2013)}]{fathi2013meningioma}
\bibinfo{author}{Fathi, A.R.}, \bibinfo{author}{Roelcke, U.}, \bibinfo{year}{2013}.
\newblock \bibinfo{title}{Meningioma}.
\newblock \bibinfo{journal}{Current neurology and neuroscience reports} \bibinfo{volume}{13}, \bibinfo{pages}{337}.
\bibitem[{Ferreira et~al.(2024)Ferreira, Luijten, Puladi, Kleesiek, Alves and Egger}]{ferreira2024brain}
\bibinfo{author}{Ferreira, A.}, \bibinfo{author}{Luijten, G.}, \bibinfo{author}{Puladi, B.}, \bibinfo{author}{Kleesiek, J.}, \bibinfo{author}{Alves, V.}, \bibinfo{author}{Egger, J.}, \bibinfo{year}{2024}.
\newblock \bibinfo{title}{Brain tumour removing and missing modality generation using 3d wdm}.
\newblock \bibinfo{journal}{arXiv preprint arXiv:2411.04630} .
\bibitem[{Goodfellow et~al.(2014)Goodfellow, Pouget-Abadie, Mirza, Xu, Warde-Farley, Ozair, Courville and Bengio}]{goodfellow2014generative}
\bibinfo{author}{Goodfellow, I.J.}, \bibinfo{author}{Pouget-Abadie, J.}, \bibinfo{author}{Mirza, M.}, \bibinfo{author}{Xu, B.}, \bibinfo{author}{Warde-Farley, D.}, \bibinfo{author}{Ozair, S.}, \bibinfo{author}{Courville, A.}, \bibinfo{author}{Bengio, Y.}, \bibinfo{year}{2014}.
\newblock \bibinfo{title}{Generative adversarial nets}.
\newblock \bibinfo{journal}{Advances in neural information processing systems} \bibinfo{volume}{27}.
\bibitem[{Guo et~al.(2023)Guo, Shi, Liu, Li, Chen, Wang, Wang, Xing, Xia, Li et~al.}]{guo2023clinical}
\bibinfo{author}{Guo, X.}, \bibinfo{author}{Shi, Y.}, \bibinfo{author}{Liu, D.}, \bibinfo{author}{Li, Y.}, \bibinfo{author}{Chen, W.}, \bibinfo{author}{Wang, Y.}, \bibinfo{author}{Wang, Y.}, \bibinfo{author}{Xing, H.}, \bibinfo{author}{Xia, Y.}, \bibinfo{author}{Li, J.}, et~al., \bibinfo{year}{2023}.
\newblock \bibinfo{title}{Clinical updates on gliomas and implications of the 5th edition of the who classification of central nervous system tumors}.
\newblock \bibinfo{journal}{Frontiers in Oncology} \bibinfo{volume}{13}, \bibinfo{pages}{1131642}.
\bibitem[{Havaei et~al.(2016)Havaei, Guizard, Chapados and Bengio}]{havaei2016hemis}
\bibinfo{author}{Havaei, M.}, \bibinfo{author}{Guizard, N.}, \bibinfo{author}{Chapados, N.}, \bibinfo{author}{Bengio, Y.}, \bibinfo{year}{2016}.
\newblock \bibinfo{title}{Hemis: Hetero-modal image segmentation}, in: \bibinfo{booktitle}{International conference on medical image computing and computer-assisted intervention}, \bibinfo{organization}{Springer}. pp. \bibinfo{pages}{469--477}.
\bibitem[{Hu et~al.(2020)Hu, Maillard, Zhang, Ciceri, La~Barbera, Bloch and Gori}]{hu2020knowledge}
\bibinfo{author}{Hu, M.}, \bibinfo{author}{Maillard, M.}, \bibinfo{author}{Zhang, Y.}, \bibinfo{author}{Ciceri, T.}, \bibinfo{author}{La~Barbera, G.}, \bibinfo{author}{Bloch, I.}, \bibinfo{author}{Gori, P.}, \bibinfo{year}{2020}.
\newblock \bibinfo{title}{Knowledge distillation from multi-modal to mono-modal segmentation networks}, in: \bibinfo{booktitle}{International Conference on Medical Image Computing and Computer-Assisted Intervention}, \bibinfo{organization}{Springer}. pp. \bibinfo{pages}{772--781}.
\bibitem[{Isensee et~al.(2021)Isensee, Jaeger, Kohl, Petersen and Maier-Hein}]{isensee2021nnu}
\bibinfo{author}{Isensee, F.}, \bibinfo{author}{Jaeger, P.F.}, \bibinfo{author}{Kohl, S.A.}, \bibinfo{author}{Petersen, J.}, \bibinfo{author}{Maier-Hein, K.H.}, \bibinfo{year}{2021}.
\newblock \bibinfo{title}{nnu-net: a self-configuring method for deep learning-based biomedical image segmentation}.
\newblock \bibinfo{journal}{Nature methods} \bibinfo{volume}{18}, \bibinfo{pages}{203--211}.
\bibitem[{Isensee et~al.(2024)Isensee, Wald, Ulrich, Baumgartner, Roy, Maier-Hein and Jaeger}]{isensee2024nnu}
\bibinfo{author}{Isensee, F.}, \bibinfo{author}{Wald, T.}, \bibinfo{author}{Ulrich, C.}, \bibinfo{author}{Baumgartner, M.}, \bibinfo{author}{Roy, S.}, \bibinfo{author}{Maier-Hein, K.}, \bibinfo{author}{Jaeger, P.F.}, \bibinfo{year}{2024}.
\newblock \bibinfo{title}{nnu-net revisited: A call for rigorous validation in 3d medical image segmentation}, in: \bibinfo{booktitle}{International Conference on Medical Image Computing and Computer-Assisted Intervention}, \bibinfo{organization}{Springer}. pp. \bibinfo{pages}{488--498}.
\bibitem[{Isola et~al.(2017)Isola, Zhu, Zhou and Efros}]{isola2017image}
\bibinfo{author}{Isola, P.}, \bibinfo{author}{Zhu, J.Y.}, \bibinfo{author}{Zhou, T.}, \bibinfo{author}{Efros, A.A.}, \bibinfo{year}{2017}.
\newblock \bibinfo{title}{Image-to-image translation with conditional adversarial networks}, in: \bibinfo{booktitle}{Proceedings of the IEEE conference on computer vision and pattern recognition}, pp. \bibinfo{pages}{1125--1134}.
\bibitem[{Kazerooni et~al.(2024)Kazerooni, Khalili, Liu, Haldar, Jiang, Anwar, Albrecht, Adewole, Anazodo, Anderson et~al.}]{kazerooni2024brain}
\bibinfo{author}{Kazerooni, A.F.}, \bibinfo{author}{Khalili, N.}, \bibinfo{author}{Liu, X.}, \bibinfo{author}{Haldar, D.}, \bibinfo{author}{Jiang, Z.}, \bibinfo{author}{Anwar, S.M.}, \bibinfo{author}{Albrecht, J.}, \bibinfo{author}{Adewole, M.}, \bibinfo{author}{Anazodo, U.}, \bibinfo{author}{Anderson, H.}, et~al., \bibinfo{year}{2024}.
\newblock \bibinfo{title}{The brain tumor segmentation (brats) challenge 2023: focus on pediatrics (cbtn-connect-dipgr-asnr-miccai brats-peds)}.
\newblock \bibinfo{journal}{ArXiv} , \bibinfo{pages}{arXiv--2305}.
\bibitem[{Kebaili et~al.(2025a)Kebaili, Lapuyade-Lahorgue, Vera and Ruan}]{kebaili2025amm}
\bibinfo{author}{Kebaili, A.}, \bibinfo{author}{Lapuyade-Lahorgue, J.}, \bibinfo{author}{Vera, P.}, \bibinfo{author}{Ruan, S.}, \bibinfo{year}{2025}a.
\newblock \bibinfo{title}{Amm-diff: Adaptive multi-modality diffusion network for missing modality imputation}, in: \bibinfo{booktitle}{2025 IEEE 22nd International Symposium on Biomedical Imaging (ISBI)}, \bibinfo{organization}{IEEE}. pp. \bibinfo{pages}{1--4}.
\bibitem[{Kebaili et~al.(2025b)Kebaili, Lapuyade-Lahorgue, Vera and Ruan}]{kebaili2025multi}
\bibinfo{author}{Kebaili, A.}, \bibinfo{author}{Lapuyade-Lahorgue, J.}, \bibinfo{author}{Vera, P.}, \bibinfo{author}{Ruan, S.}, \bibinfo{year}{2025}b.
\newblock \bibinfo{title}{Multi-modal mri synthesis with conditional latent diffusion models for data augmentation in tumor segmentation}.
\newblock \bibinfo{journal}{Computerized Medical Imaging and Graphics} \bibinfo{volume}{123}, \bibinfo{pages}{102532}.
\bibitem[{Kim and Park(2024)}]{kim2024adaptive}
\bibinfo{author}{Kim, J.}, \bibinfo{author}{Park, H.}, \bibinfo{year}{2024}.
\newblock \bibinfo{title}{Adaptive latent diffusion model for 3d medical image to image translation: Multi-modal magnetic resonance imaging study}, in: \bibinfo{booktitle}{Proceedings of the IEEE/CVF Winter conference on applications of computer Vision}, pp. \bibinfo{pages}{7604--7613}.
\bibitem[{LaBella et~al.(2023)LaBella, Adewole, Alonso-Basanta, Altes, Anwar, Baid, Bergquist, Bhalerao, Chen, Chung et~al.}]{labella2023asnr}
\bibinfo{author}{LaBella, D.}, \bibinfo{author}{Adewole, M.}, \bibinfo{author}{Alonso-Basanta, M.}, \bibinfo{author}{Altes, T.}, \bibinfo{author}{Anwar, S.M.}, \bibinfo{author}{Baid, U.}, \bibinfo{author}{Bergquist, T.}, \bibinfo{author}{Bhalerao, R.}, \bibinfo{author}{Chen, S.}, \bibinfo{author}{Chung, V.}, et~al., \bibinfo{year}{2023}.
\newblock \bibinfo{title}{The asnr-miccai brain tumor segmentation (brats) challenge 2023: Intracranial meningioma}.
\newblock \bibinfo{journal}{arXiv preprint arXiv:2305.07642} .
\bibitem[{Li et~al.(2024)Li, Conte, Hu, Anwar, Kofler, Ezhov, van Leemput, Piraud, Diaz, Cole et~al.}]{li2024brain}
\bibinfo{author}{Li, H.B.}, \bibinfo{author}{Conte, G.M.}, \bibinfo{author}{Hu, Q.}, \bibinfo{author}{Anwar, S.M.}, \bibinfo{author}{Kofler, F.}, \bibinfo{author}{Ezhov, I.}, \bibinfo{author}{van Leemput, K.}, \bibinfo{author}{Piraud, M.}, \bibinfo{author}{Diaz, M.}, \bibinfo{author}{Cole, B.}, et~al., \bibinfo{year}{2024}.
\newblock \bibinfo{title}{The brain tumor segmentation (brats) challenge 2023: Brain mr image synthesis for tumor segmentation (brasyn)}.
\newblock \bibinfo{journal}{ArXiv} , \bibinfo{pages}{arXiv--2305}.
\bibitem[{Liu et~al.(2023)Liu, Wei, Li and Zhou}]{liu2023sfusion}
\bibinfo{author}{Liu, Z.}, \bibinfo{author}{Wei, J.}, \bibinfo{author}{Li, R.}, \bibinfo{author}{Zhou, J.}, \bibinfo{year}{2023}.
\newblock \bibinfo{title}{Sfusion: Self-attention based n-to-one multimodal fusion block}, in: \bibinfo{booktitle}{International conference on medical image computing and computer-assisted intervention}, \bibinfo{organization}{Springer}. pp. \bibinfo{pages}{159--169}.
\bibitem[{Longuefosse et~al.(2026)Longuefosse, Gonzalez, Baldacci and Seita}]{longuefosse2026intensity}
\bibinfo{author}{Longuefosse, A.}, \bibinfo{author}{Gonzalez, J.S.}, \bibinfo{author}{Baldacci, F.}, \bibinfo{author}{Seita, J.}, \bibinfo{year}{2026}.
\newblock \bibinfo{title}{From intensity to anatomy: Optimizing nnu-net for mr-to-ct translation on synthrad2025}, in: \bibinfo{booktitle}{2026 IEEE 23rd International Symposium on Biomedical Imaging (ISBI)}, \bibinfo{organization}{IEEE}. pp. \bibinfo{pages}{1--4}.
\bibitem[{Longuefosse et~al.(2025)Longuefosse, Denis~de Senneville, Dournes, Benlala, Baldacci and Desbarats}]{longuefosse2025anatomical}
\bibinfo{author}{Longuefosse, A.}, \bibinfo{author}{Denis~de Senneville, B.}, \bibinfo{author}{Dournes, G.}, \bibinfo{author}{Benlala, I.}, \bibinfo{author}{Baldacci, F.}, \bibinfo{author}{Desbarats, P.}, \bibinfo{year}{2025}.
\newblock \bibinfo{title}{Anatomical feature-prioritized loss for enhanced mr to ct translation}.
\newblock \bibinfo{journal}{Physics in Medicine \& Biology} \bibinfo{volume}{70}, \bibinfo{pages}{145012}.
\bibitem[{Longuefosse et~al.(2024)}]{longuefosse2024adapted}
\bibinfo{author}{Longuefosse, A.}, et~al., \bibinfo{year}{2024}.
\newblock \bibinfo{title}{Adapted nnu-net: A robust baseline for cross-modality synthesis and medical image inpainting}, in: \bibinfo{booktitle}{Simulation and Synthesis in Medical Imaging: 8th International Workshop, SASHIMI 2023, Held in Conjunction with MICCAI 2023, Vancouver, BC, Canada, October 8, 2023, Proceedings}, \bibinfo{publisher}{Springer}. pp. \bibinfo{pages}{24--33}.
\bibitem[{Meng et~al.(2024)Meng, Sun, Xu, He and Shen}]{meng2024multi}
\bibinfo{author}{Meng, X.}, \bibinfo{author}{Sun, K.}, \bibinfo{author}{Xu, J.}, \bibinfo{author}{He, X.}, \bibinfo{author}{Shen, D.}, \bibinfo{year}{2024}.
\newblock \bibinfo{title}{Multi-modal modality-masked diffusion network for brain mri synthesis with random modality missing}.
\newblock \bibinfo{journal}{IEEE Transactions on Medical Imaging} \bibinfo{volume}{43}, \bibinfo{pages}{2587--2598}.
\bibitem[{Menze et~al.(2014)Menze, Jakab, Bauer, Kalpathy-Cramer, Farahani, Kirby, Burren, Porz, Slotboom, Wiest et~al.}]{menze2014multimodal}
\bibinfo{author}{Menze, B.H.}, \bibinfo{author}{Jakab, A.}, \bibinfo{author}{Bauer, S.}, \bibinfo{author}{Kalpathy-Cramer, J.}, \bibinfo{author}{Farahani, K.}, \bibinfo{author}{Kirby, J.}, \bibinfo{author}{Burren, Y.}, \bibinfo{author}{Porz, N.}, \bibinfo{author}{Slotboom, J.}, \bibinfo{author}{Wiest, R.}, et~al., \bibinfo{year}{2014}.
\newblock \bibinfo{title}{The multimodal brain tumor image segmentation benchmark (brats)}.
\newblock \bibinfo{journal}{IEEE transactions on medical imaging} \bibinfo{volume}{34}, \bibinfo{pages}{1993--2024}.
\bibitem[{Nie et~al.(2018)Nie, Trullo, Lian, Wang, Petitjean, Ruan, Wang and Shen}]{nie2018medical}
\bibinfo{author}{Nie, D.}, \bibinfo{author}{Trullo, R.}, \bibinfo{author}{Lian, J.}, \bibinfo{author}{Wang, L.}, \bibinfo{author}{Petitjean, C.}, \bibinfo{author}{Ruan, S.}, \bibinfo{author}{Wang, Q.}, \bibinfo{author}{Shen, D.}, \bibinfo{year}{2018}.
\newblock \bibinfo{title}{Medical image synthesis with deep convolutional adversarial networks}.
\newblock \bibinfo{journal}{IEEE Transactions on Biomedical Engineering} \bibinfo{volume}{65}, \bibinfo{pages}{2720--2730}.
\bibitem[{Pipoli et~al.(2025)Pipoli, Saporita, Marchesini, Grana, Ficarra and Bolelli}]{pipoli2025fuse}
\bibinfo{author}{Pipoli, V.}, \bibinfo{author}{Saporita, A.}, \bibinfo{author}{Marchesini, K.}, \bibinfo{author}{Grana, C.}, \bibinfo{author}{Ficarra, E.}, \bibinfo{author}{Bolelli, F.}, \bibinfo{year}{2025}.
\newblock \bibinfo{title}{Im-fuse: A mamba-based fusion block for brain tumor segmentation with incomplete modalities}, in: \bibinfo{booktitle}{International Conference on Medical Image Computing and Computer-Assisted Intervention}, \bibinfo{organization}{Springer}. pp. \bibinfo{pages}{225--235}.
\bibitem[{Sharma and Hamarneh(2019)}]{sharma2019missing}
\bibinfo{author}{Sharma, A.}, \bibinfo{author}{Hamarneh, G.}, \bibinfo{year}{2019}.
\newblock \bibinfo{title}{Missing mri pulse sequence synthesis using multi-modal generative adversarial network}.
\newblock \bibinfo{journal}{IEEE transactions on medical imaging} \bibinfo{volume}{39}, \bibinfo{pages}{1170--1183}.
\bibitem[{Shi(2024)}]{shi2024hyperparameters}
\bibinfo{author}{Shi, B.}, \bibinfo{year}{2024}.
\newblock \bibinfo{title}{On the hyperparameters in stochastic gradient descent with momentum}.
\newblock \bibinfo{journal}{Journal of Machine Learning Research} \bibinfo{volume}{25}, \bibinfo{pages}{1--40}.
\bibitem[{Taha and Hanbury(2015)}]{taha2015metrics}
\bibinfo{author}{Taha, A.A.}, \bibinfo{author}{Hanbury, A.}, \bibinfo{year}{2015}.
\newblock \bibinfo{title}{Metrics for evaluating 3d medical image segmentation: analysis, selection, and tool}.
\newblock \bibinfo{journal}{BMC medical imaging} \bibinfo{volume}{15}, \bibinfo{pages}{29}.
\bibitem[{Tavse et~al.(2022)Tavse, Varadarajan, Bachute, Gite and Kotecha}]{tavse2022systematic}
\bibinfo{author}{Tavse, S.}, \bibinfo{author}{Varadarajan, V.}, \bibinfo{author}{Bachute, M.}, \bibinfo{author}{Gite, S.}, \bibinfo{author}{Kotecha, K.}, \bibinfo{year}{2022}.
\newblock \bibinfo{title}{A systematic literature review on applications of gan-synthesized images for brain mri}.
\newblock \bibinfo{journal}{Future Internet} \bibinfo{volume}{14}, \bibinfo{pages}{351}.
\bibitem[{Weller et~al.(2015)Weller, Wick, Aldape, Brada, Berger, Pfister, Nishikawa, Rosenthal, Wen, Stupp et~al.}]{weller2015glioma}
\bibinfo{author}{Weller, M.}, \bibinfo{author}{Wick, W.}, \bibinfo{author}{Aldape, K.}, \bibinfo{author}{Brada, M.}, \bibinfo{author}{Berger, M.}, \bibinfo{author}{Pfister, S.M.}, \bibinfo{author}{Nishikawa, R.}, \bibinfo{author}{Rosenthal, M.}, \bibinfo{author}{Wen, P.Y.}, \bibinfo{author}{Stupp, R.}, et~al., \bibinfo{year}{2015}.
\newblock \bibinfo{title}{Glioma}.
\newblock \bibinfo{journal}{Nature reviews Disease primers} \bibinfo{volume}{1}, \bibinfo{pages}{1--18}.
\bibitem[{Yu et~al.(2019)Yu, Zhou, Wang, Shi, Fripp and Bourgeat}]{yu2019ea}
\bibinfo{author}{Yu, B.}, \bibinfo{author}{Zhou, L.}, \bibinfo{author}{Wang, L.}, \bibinfo{author}{Shi, Y.}, \bibinfo{author}{Fripp, J.}, \bibinfo{author}{Bourgeat, P.}, \bibinfo{year}{2019}.
\newblock \bibinfo{title}{Ea-gans: edge-aware generative adversarial networks for cross-modality mr image synthesis}.
\newblock \bibinfo{journal}{IEEE transactions on medical imaging} \bibinfo{volume}{38}, \bibinfo{pages}{1750--1762}.
\bibitem[{Zhang et~al.(2022)Zhang, He, Yang, Li, Wei, Huang, Zhang, He and Zheng}]{zhang2022mmformer}
\bibinfo{author}{Zhang, Y.}, \bibinfo{author}{He, N.}, \bibinfo{author}{Yang, J.}, \bibinfo{author}{Li, Y.}, \bibinfo{author}{Wei, D.}, \bibinfo{author}{Huang, Y.}, \bibinfo{author}{Zhang, Y.}, \bibinfo{author}{He, Z.}, \bibinfo{author}{Zheng, Y.}, \bibinfo{year}{2022}.
\newblock \bibinfo{title}{mmformer: Multimodal medical transformer for incomplete multimodal learning of brain tumor segmentation}, in: \bibinfo{booktitle}{International conference on medical image computing and computer-assisted intervention}, \bibinfo{organization}{Springer}. pp. \bibinfo{pages}{107--117}.
\bibitem[{Zhang et~al.(2024)Zhang, Peng, Wang, Song, Li and Zhou}]{zhang2024unified}
\bibinfo{author}{Zhang, Y.}, \bibinfo{author}{Peng, C.}, \bibinfo{author}{Wang, Q.}, \bibinfo{author}{Song, D.}, \bibinfo{author}{Li, K.}, \bibinfo{author}{Zhou, S.K.}, \bibinfo{year}{2024}.
\newblock \bibinfo{title}{Unified multi-modal image synthesis for missing modality imputation}.
\newblock \bibinfo{journal}{IEEE Transactions on Medical Imaging} \bibinfo{volume}{44}, \bibinfo{pages}{4--18}.
\bibitem[{Zhu et~al.(2017)Zhu, Park, Isola and Efros}]{zhu2017unpaired}
\bibinfo{author}{Zhu, J.Y.}, \bibinfo{author}{Park, T.}, \bibinfo{author}{Isola, P.}, \bibinfo{author}{Efros, A.A.}, \bibinfo{year}{2017}.
\newblock \bibinfo{title}{Unpaired image-to-image translation using cycle-consistent adversarial networks}, in: \bibinfo{booktitle}{Proceedings of the IEEE international conference on computer vision}, pp. \bibinfo{pages}{2223--2232}.
\bibitem[{Zhu et~al.(2026)Zhu, Chen, Chen, Jiang, Zhou, Wang, Qin, Wang and Tian}]{zhu2026no}
\bibinfo{author}{Zhu, S.}, \bibinfo{author}{Chen, Y.}, \bibinfo{author}{Chen, W.}, \bibinfo{author}{Jiang, S.}, \bibinfo{author}{Zhou, G.}, \bibinfo{author}{Wang, Y.}, \bibinfo{author}{Qin, F.}, \bibinfo{author}{Wang, C.}, \bibinfo{author}{Tian, Q.}, \bibinfo{year}{2026}.
\newblock \bibinfo{title}{No modality left behind: Adapting to missing modalities via knowledge distillation for brain tumor segmentation}.
\newblock \bibinfo{journal}{Medical Image Analysis} , \bibinfo{pages}{104108}.

\end{thebibliography}
\end{document}